\documentclass[11pt,twocolumn,aps,prb,superscriptaddress,reprint,citeautoscript,sort&compress,square]{revtex4-1}

\usepackage{graphicx,amsmath,amssymb,color}
\usepackage{tabularx, ctable}
\usepackage{ulem}

\newcommand{\EM}[1]{\textcolor{black}{#1}}

\newcommand{\SI}[1]{\textcolor{black}{#1}}

\newcommand{\IYP}[1]{\textcolor{blue}{#1}}

\usepackage[breaklinks=true,colorlinks=false,citecolor=blue,linkcolor=blue,urlcolor=blue] {hyperref}

\setcitestyle{super}

\renewcommand*{\Im}{\operatorname{Im}}
\renewcommand*{\Re}{\operatorname{Re}}

\begin{document}


\title{Cyclotron resonance overtones and near-field magnetoabsorption via terahertz Bernstein modes in graphene}

\author{D. A. Bandurin$^{*,+}$}
\affiliation{Department of Physics, Massachusetts Institute of Technology, Cambridge, Massachusetts 02139}

\author{E. M\"onch$^+$}
\affiliation{Terahertz Center, University of
Regensburg, 93040 Regensburg, Germany}
 
\author{K. Kapralov}
\affiliation{Center for Photonics and 2d Materials, Moscow Institute of Physics and Technology (National Research University), Dolgoprudny 141700, Russia}
 
\author{I. Y. Phinney}
\affiliation{Department of Physics, Massachusetts Institute of Technology, Cambridge, Massachusetts 02139}
 
\author{K. Lindner}
\affiliation{Terahertz Center, University of
Regensburg, 93040 Regensburg, Germany}

\author{S. Liu}
\affiliation{The Tim Taylor Department of Chemical Engineering, Kansas State University, Manhattan, KS, 66506, USA.}

\author{J. H. Edgar}
\affiliation{The Tim Taylor Department of Chemical Engineering, Kansas State University, Manhattan, KS, 66506, USA.}

\author{I. A. Dmitriev}
\affiliation{Terahertz Center, University of
Regensburg, 93040 Regensburg, Germany}

\author{P. Jarillo-Herrero}
\affiliation{Department of Physics, Massachusetts Institute of Technology, Cambridge, Massachusetts 02139}

\author{D. Svintsov*}
\affiliation{Center for Photonics and 2d Materials, Moscow Institute of Physics and Technology (National Research University), Dolgoprudny 141700, Russia}
 
\author{S.D. Ganichev*}
 \affiliation{Terahertz Center, University of
Regensburg, 93040 Regensburg, Germany}

\begin{abstract}
\textbf{
Two-dimensional electron systems subjected to a perpendicular magnetic field absorb electromagnetic radiation via the cyclotron resonance (CR).
Here we report a qualitative breach of this well-known behaviour in graphene. Our study of the terahertz photoresponse reveals a resonant burst at the main overtone of the CR, drastically exceeding the signal detected at the position of the ordinary CR. In accordance with the developed theory, the photoresponse dependencies on the magnetic field, doping level, and sample geometry suggest that the origin of this anomaly lies in the near-field magnetoabsorption facilitated by the Bernstein modes, ultra-slow magnetoplasmonic excitations reshaped by nonlocal electron dynamics. 
Close to the CR harmonics, these modes are characterized by a flat dispersion and a diverging plasmonic density of states that strongly amplifies the radiation absorption. Besides fundamental interest, our experimental results and developed theory show that the radiation absorption via nonlocal collective modes can facilitate a strong photoresponse, a behaviour potentially useful for infrared and terahertz technology.}

\begin{center}
$^{+}$ Equally contributed authors. 
\end{center}

\end{abstract}

\maketitle

Understanding light-matter interaction at the nanoscale is a fundamental challenge at the boundary between condensed-matter physics and nanophotonics~\cite{novotny2012principles}.
At subwavelength scales, electromagnetic (EM) fields tend to be bound to interfaces and inhomogeneities in the material system  via various collective excitations~\cite{Basovaag1992}. A prominent example of such excitations is plasmons, which represent coupled oscillations of charge carriers
and an associated EM field. Recent years have seen a renaissance in the field of plasmonics brought upon by the discovery of clean, two-dimensional electron systems (2DES)~\cite{Muravev_PRL_DrasticReduction,Faist_PRL_UltrastrongCoupling} -- most prominently, graphene, with its long-lived gate-tunable plasmons~\cite{ni2018fundamental,Koppens_2015_ConfinedLowLoss}. These material platforms enable field compression by two orders of magnitude in all directions as compared to free-space radiation~\cite{ni2018fundamental}, thus opening an avenue for the exploration of
various quantum electrodynamic phenomena~\cite{Koppens_2021_QuantNanophotonicsWith2DM,QNonlocal} and the development of practical infrared and THz devices~\cite{Xia2014,Muravev_plasmonic_Detector,LundebergTherm,Bandurin2018}.


\begin{figure*}[ht!]
	\centering\includegraphics[width=\linewidth]{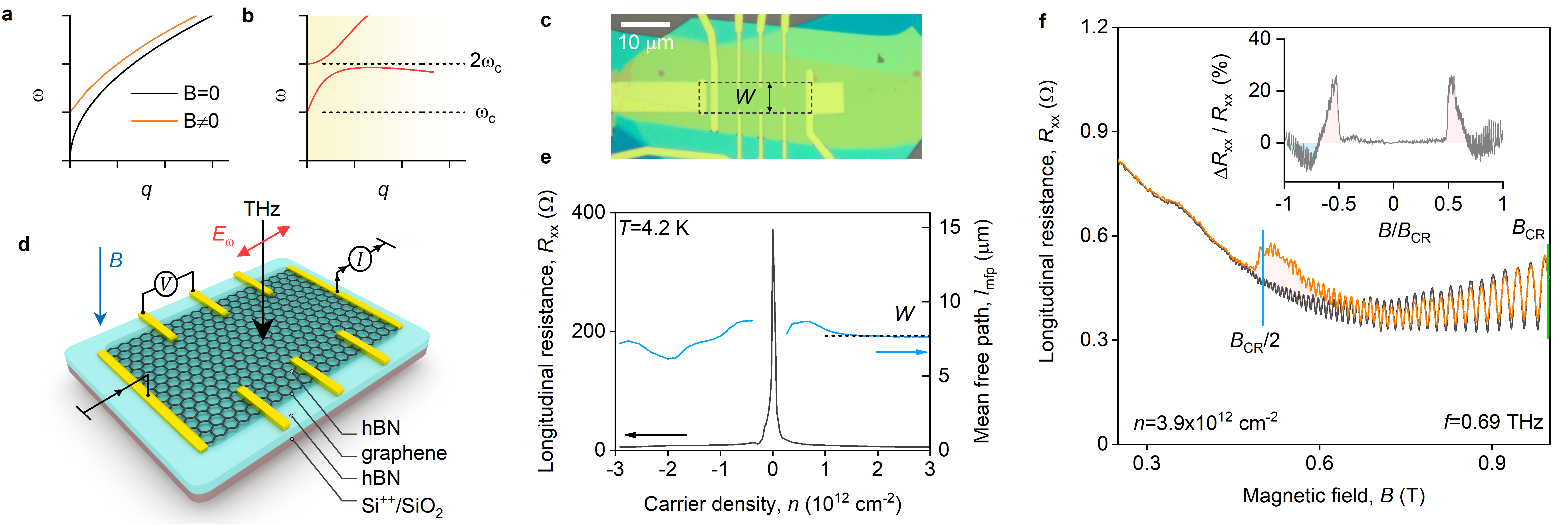}
	\caption{\textbf{Anomalous THz photoresponse of doped graphene.} \textbf{a,} The dispersion of plasmons in a 2DES at zero and finite $B$. \textbf{b,} The dispersion of BMs. At large $q\gtrsim1/R_\mathrm{c}$ the magnetoplasmon dispersion splits into a series of branches with anticrossing regions in the vicinity of the cyclotron frequency harmonics.
	\textbf{ c,} Optical photograph of an encapsulated graphene device. Yellow leads are gold contacts to graphene. Green corresponds to hBN.  \textbf{d,} Measurement configuration: THz radiation is incident upon doped graphene subjected to perpendicular magnetic field. The photovoltage or photoresistance is measured between different pairs of contacts. \textbf{e,} The longitudinal resistance, $R_\mathrm{xx}$, of the device shown in (c) as a function of carrier density at 4.2~K and $B=0$, together with the 
	corresponding mean free path, $l_\mathrm{mfp}$, obtained using the Drude model. \textbf{f,} Device resistance $R_\mathrm{xx}$ as a function of $B$ measured in the dark (black curve) and in the presence of 0.69 THz radiation. Green and blue vertical lines correspond to the magnetic fields at which the CR and its second harmonics are expected at given $n$.  Inset: Photoresistance $\Delta R_\mathrm{xx}$ normalized to $R_\mathrm{xx}$ as a function $B/B_\mathrm{CR}$. $T=4.2$~K.   }
	\label{Fig_1}
\end{figure*}

The mechanisms of light-matter interaction at even smaller spatial scales (in the deep subwavelength regime) 
are more intricate, and are generally governed by 
the nonlocality of a material's current-field relation, parameterized by its nonlocal conductivity~\cite{Ciraci-2012-Limits_of_Plasmonic_Enhancement}.
On the one hand, the nonlocal conductivity carries valuable information about quasiparticle dynamics~\cite{Gonccalves-2020-quantum} and interactions~\cite{Lundeberg2017} inaccessible from measurements at larger spatial scales. On the other hand, the nonlocality is expected to smear the electric fields at the smallest scales~\cite{Pendry-2017-Compacted_dimensions}, and thereby set the ultimate limits to field confinement and slowing of light in nanophotonics~\cite{Iranzo_2018_ConfinementLimits}.
However, despite fundamental interest and practical importance, accessing the nonlocal regime of light-matter interaction has been exceedingly difficult~\cite{Heitman_1996_FermiPressure,Lundeberg2017,Iranzo_2018_ConfinementLimits,NonlocalLEES}.
This has stimulated the belief that nonlocal effects yield only corrections to the predictions of the local conductivity model~\cite{Heitman_1996_FermiPressure,Lundeberg2017} and therefore are routinely disregarded.

In this Article, we show that the opposite is in fact true and demonstrate a prominent signature of nonlocal light-matter interaction in a high-quality graphene device exposed to THz radiation and subjected to a classically strong magnetic field, $B$. The $B$ field modifies the dispersion of the two-dimensional plasmons which, at zero wave vector, $q$,  acquires a gap (Fig.~\ref{Fig_1}a) below the frequency of the electrons' cyclotron motion, $\omega_\mathrm{c}$~\cite{Magnetoplasmons-graphene-1,Magnetoplasmons-graphene-2}. Moreover, at large $q\gtrsim 1/R_\mathrm{c}$, where $R_\mathrm{c}$ is the cyclotron radius, and\IYP{,} in sufficiently clean 2DES, the dispersion splits into a series of branches (Fig.~\ref{Fig_1}b), an intricate behaviour stemming from the nonlocality of the material's conductivity~\cite{Quinn_1974_Bernstein2DES} (see below). 
The emergent collective excitations are dubbed Bernstein modes~\cite{Bernstein_1958} (BMs) and  were first theorized to exist in the field of plasma physics many decades ago~\cite{Sitenko1957,Bernstein_1958} and later explored in 2DES~\cite{Quinn_1974_Bernstein2DES,BMBatke,Gudmundsson_1995_Bernstein_modes_wires_dots,BMBatke2,BM_photoconductivity,Bernstein-modes-graphene}. 
We show that the vanishing group velocity of the BMs and large plasmonic density of states~\cite{Volkov2014} facilitate strong magnetoabsorption in graphene devices in the vicinity of the CR overtones. We detect it as a giant photoresistance peak, exceeding the signal due to the ordinary CR by few orders of magnitude. This observation is in striking contrast with the commonly accepted scenario in which, with few exceptions~\cite{X2-peak-1,X-2_Zudov}, resonant absorption at multiple CR harmonics is a weaker effect appearing only due to non-uniform fields~\cite{Chaplik_1985_CR_overtones,Bangert1996,Umansky}. Our findings reveal the importance of nonlocal light-matter interaction in subwavelength THz devices.

\textbf{Design and characterization of the graphene plasmonic device. } Our sample is a multi-terminal device made of graphene encapsulated between hexagonal boron nitride (hBN) crystals that provides the best environment for a high-mobility electron transport. The device was fabricated using a hot transfer technique described elsewhere~\cite{Purdie2018} (See Methods) and assembled on top of a Si/SiO$_2$ substrate that acts as gate controlling the carrier density, $n$ (Fig.~\ref{Fig_1}c). The device shape was defined by the shape of the underlying graphene flake which for this sample was about $8~\mu$m wide and more than $30~\mu$m long (dashed contour in Fig.~\ref{Fig_1}c). To ensure the superior quality of the obtained heterostructure, we refrained from its etching into a standard Hall bar structure, but instead, left natural graphene edge intact. The contacts were then embedded into the device channel by a combination of selective ion etching and evaporation of thin metal layer (See Methods). This configuration is particularly suitable for plasmonic experiments in which incident radiation launches plasma waves via scattering by narrow metal contacts thus mitigating the momentum mismatch between incident THz photons and graphene plasmons~\cite{Alonso-Gonzlez-2014-spatial_conductivity_patterns}.  The response of our devices to incident THz radiation was recorded via photoresistance and photovoltage measurements, both performed in Faraday configuration with the laser beam and magnetic field oriented perpendicular to the graphene plane (see Fig.~\ref{Fig_1}d and Methods). 


Figure~\ref{Fig_1}e presents the longitudnial resistance, $R_\mathrm{xx}$, of our device which exhibits a standard graphene behavior: $R_\mathrm{xx}$ peaks at the charge neutrality point (CNP) and drops rapidly upon doping. Invoking the Drude model we estimated the mean free path, $l_\mathrm{mfp}$, of our device at $B=0$~T (blue curve in Fig.~\ref{Fig_1}e). Away from the CNP, it is comparable to the the device width, $W=8~\mu$m, indicating micrometer-scale ballistic transport (\SI{Supplementary Section 1}). 
At the same time, the magnetoresistance,  $R_\mathrm{xx}(B)$, presented in \SI{Supplementary Section 1} demonstrates the transition to a diffusive transport regime at $B\gtrsim 0.5$ T. This transition is revealed by the observation of phonon-induced resistance oscillations~\cite{zudovPIRO2001,raichevPIRO2009,RoshMP}, as well as their complex transformation under strong dc bias which reflects the tilt of Landau levels by the Hall electric field.~\cite{ZhangPHIRO2008,DmitrievPHIRO2010} 

\begin{figure*}[ht!]
	\centering\includegraphics[width=1\linewidth]{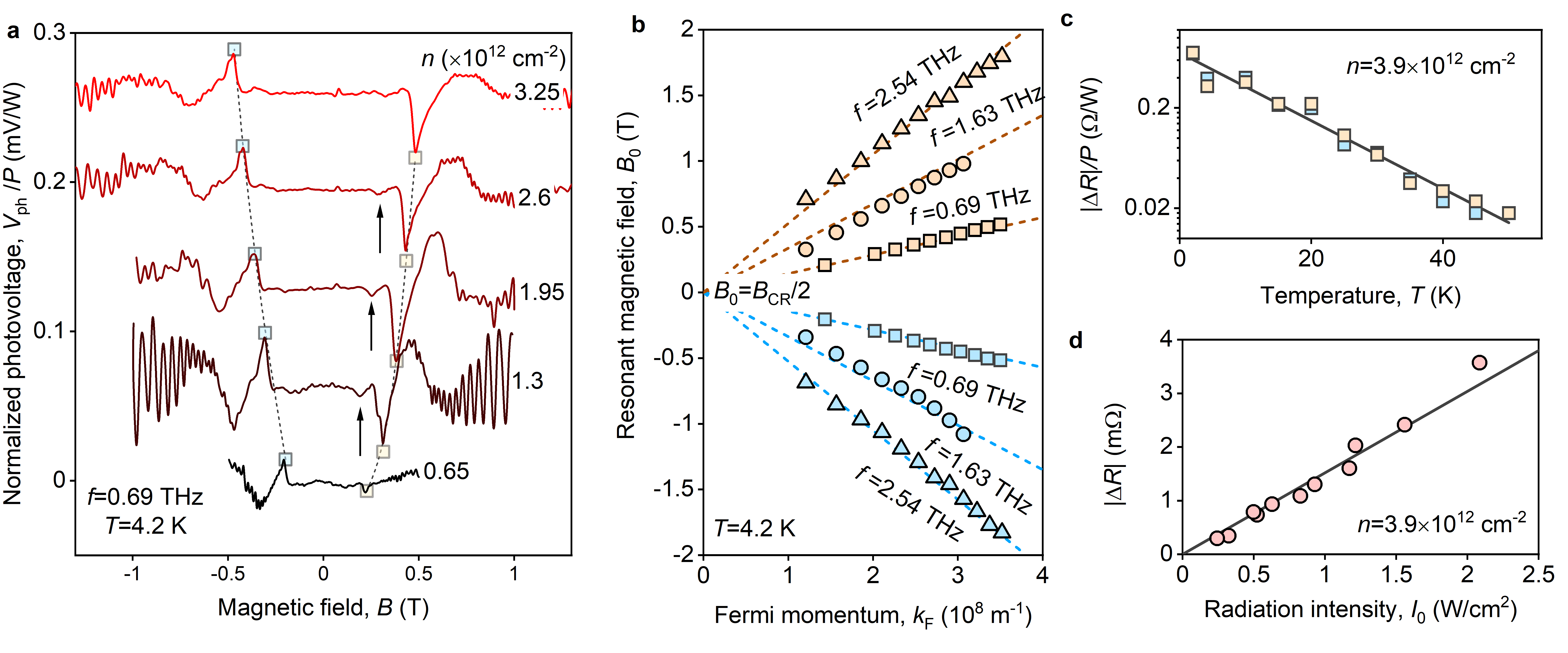}
	\caption{\textbf{CR overtones in photovoltage and photoresistance.} \textbf{a,} $V_\mathrm{ph}$ as a function of $B$ for varying $V_\mathrm{g}$ measured in response to $f=0.69$~THz radiation. Circles (arrows) mark features near the main (higher order) CR overtones. $T=4.2~$K. \textbf{b,} Resonant magnetic field $B_\mathrm{0}$ of the major photovoltage (or photoresistance) peaks  plotted against $k_\mathrm{F}$ for varying $f$ of incoming radiation (symbols). Dashed lines: calculated positions of the main CR overtone $B_\textrm{CR}/2$ as a function of $k_\mathrm{F}$ for given $f$. \textbf{c,} Peak height of the resonant photoresistance as a function of $T$ measured at $f=0.69~$THz. Solid line: Fit to $|\Delta R|\propto\exp(-T/T_\mathrm{0})$ with $T_\mathrm{0}=12.5~$K. \textbf{d,} Example of the beam intensity dependence of the resonant photoresistance recorded in response to $f=2.54~$THz radiation at given $n$. Solid line: Linear fit to $|\Delta R|\propto I_0$.
	}
	\label{Fig2}
\end{figure*}


\textbf{THz-driven magnetoresistance.} 
Figure~\ref{Fig_1}f shows typical changes in the $B$-dependence of the longitudinal resistance, $R_\mathrm{xx}$, occurring in our device in the presence of THz radiation. 
In the dark (black curve), the device features negative magnetoresistance for $|B|<0.5$~T, which is followed by the Shubnikov-de Haas oscillations (SdHO) at $|B|>0.5$~T, a standard behaviour for high-quality graphene devices~\cite{MR_graphene} (See \SI{ Supplementary Section~1} for the full magnetoresistance data). When the 0.69~THz radiation is turned on, the low-B behaviour of $R_\mathrm{xx}$ remains intact, whereas the  SdHO amplitude is slightly suppressed, presumably because of the radiation-induced increase in the electronic temperature to which SdHO are highly sensitive. A salient feature of the $R_\mathrm{xx}(B)$ dependence measured in the presence of radiation is a pronounced resistivity spike emerging at $B\sim0.5$~T (see \SI{Supplementary Section 2} for further examples). The inset of Fig.~\ref{Fig_1}f shows the difference between the orange and black curves, $\Delta R$, that isolates this anomalous feature from the magnetoresistance background and is later referred to as photoresistance.
The strength of the effect is clearly revealed when $\Delta R$ is further normalized by $R_\mathrm{xx}$. Strikingly, the anomalous part of the photoresponse peaks at the magnetic field  close to $B_\mathrm{CR}/2$, where 
$B_\mathrm{CR}$ is the magnetic field at which the CR is expected. Moreover, the $\Delta R$ peak is highly asymmetric with  respect to $B_0$, namely: $\Delta R$ spikes rapidly upon approaching $B_0$ from the left and exhibits a long tail on the right hand side from $B_0$ (see below).
Notably, the CR itself, while being allegedly stronger effect, was absent in the data shown in Fig.~\ref{Fig_1}f and only revealed itself at smaller $n$ close to the CNP or at higher $f$ (\SI{Supplementary Section 3}).

\textbf{Characterization of the photoresponse close to the CR overtones.} Figure~\ref{Fig2}a details our observations further by showing yet another signature of the anomalous photoresponse in the same device, but, in this case, in the photovoltage $V_\mathrm{ph}$ dependences on $B$ (See Methods for the measurements details). Similarly to the case of $\Delta R(B)$ data, we observed pronounced asymmetric peaks in $V_\mathrm{ph}$ for all gate voltages, $V_\mathrm{g}$. In Fig.~\ref{Fig2}b we plot the experimentally determined peak positions $B_0$ versus the Fermi momentum, $k_\mathrm{F}=\sqrt{\pi n}$, to demonstrate their linear relation. For all $f$, we find that the latter closely follows the $B_\mathrm{CR}/2=\pi f \hbar k_\mathrm{F}/e v_\mathrm{F}$
 dependence with graphene Fermi velocity, $v_\mathrm{F}=10^6~$m/s, reflecting that the observed photoresponse anomaly emerges close to the main CR overtone, $B_\mathrm{0}\approx B_\mathrm{CR}/2$. We further note, that similarly asymmetric yet weaker peaks in $V_\mathrm{ph}(B)$ dependencies were also observed in the vicinity of the higher order CR overtones, see Figs.~\ref{Fig2}a and \SI{S4}.

With increasing temperature $T$, the magnitude $\Delta R$ of the resonant photoresponse at $B$ near $B_\mathrm{CR}/2$ drops, see Fig.~\ref{Fig2}c.
This drop follows approximately an exponential decay law $\exp(- T/T_\mathrm{0})$ shown in Fig.~\ref{Fig2}c with 
the characteristic temperature \SI{$T_\mathrm{0}\approx12.5~K$} for $f=0.69~$THz. 
The maximum $T$ at which the photoresistance spike was detected at all $n$ and $f$ did not exceed 50 K above which the effect was fully suppressed (See \SI{Supplementary Section~4} for the detailed $T$~dependence). Figure~\ref{Fig2}d shows the peak height $|\Delta R|$ at $B=B_0$ measured at $T=4.2~$K
as a function of the laser beam intensity, $I_\mathrm{0}$, and reveals  
their direct proportionality at low $I$, $|\Delta R|\propto I $,  with a very slight 
tendency to a superlinear behaviour at $I_\mathrm{0}>1.5~$W/cm$^2$ (See \SI{Supplementary Section~4} for the detailed $I_\mathrm{0}$~dependence). This threshold-free linear behaviour excludes any instabilities-related mechanisms of the observed photoresponse~\cite{Volkov2014} (see below).


We also explored the polarization dependence of the photoresponse anomalies at the CR overtones and found that close to $B=B_\mathrm{CR}/2$, the signal was neither sensitive to the polarization angle of the linearly-polarized beam nor it was affected by the radiation helicity as we show in \SI{Supplementary Section~5}. This is in strong contrast to the
conventional behavior of the ordinary CR in which the handedness of the radiation's polarization
defines the magnetic field direction at which the resonant magnetoabsorption occurs. 

\begin{figure*}[ht!]
	\centering\includegraphics[width=0.9\linewidth]{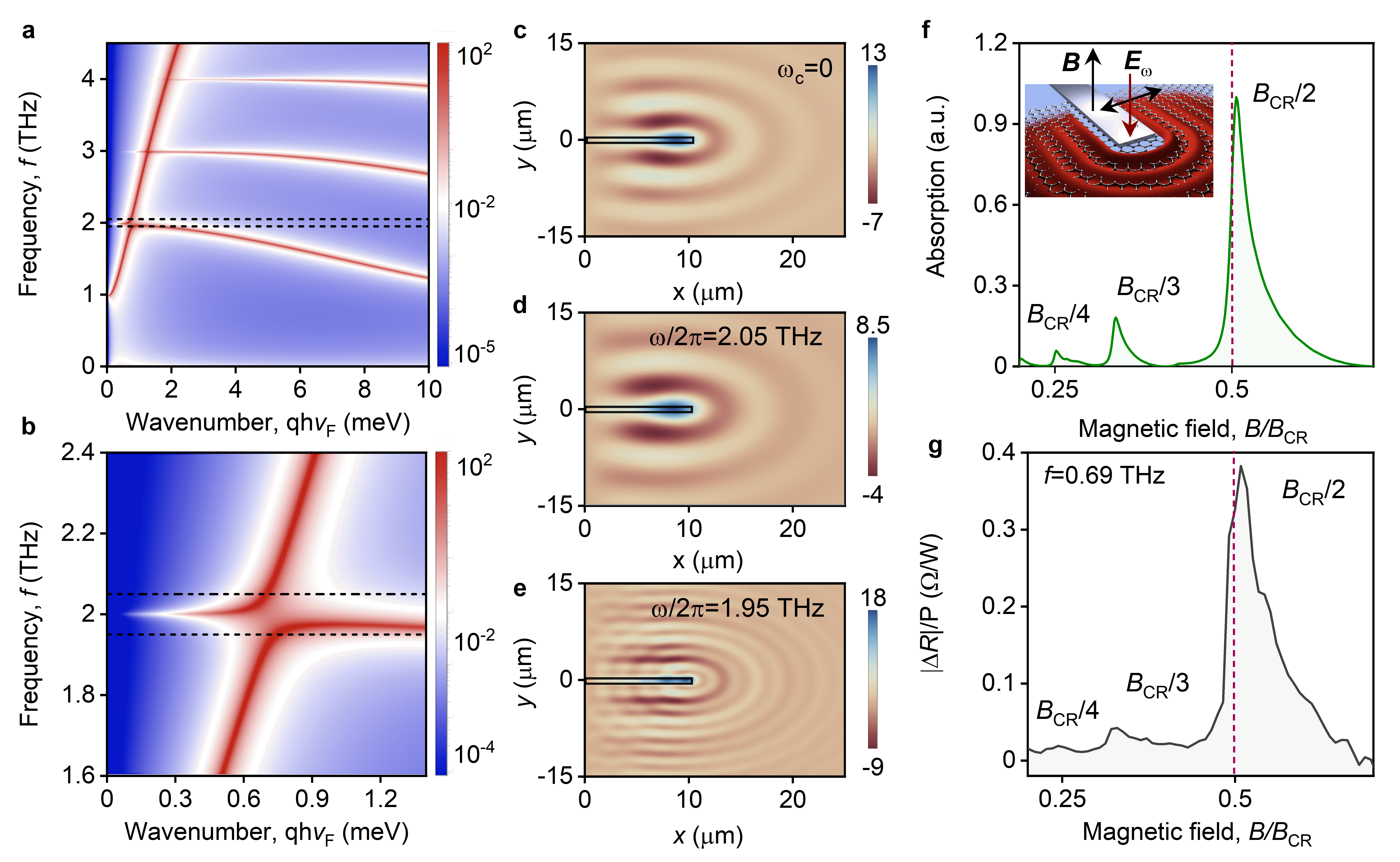}
	\caption{\textbf{Nonlocal THz magnetoabsorption in graphene.} \textbf{a,} The loss function $\Im \varepsilon^{-1}({\bf q},\omega)$ in the presence of magnetic field calculated for $\omega_\mathrm{c}/2\pi=1~$THz with perfectly conducting metal gate. Dashed horizontal lines correspond to the  frequencies $\omega/2\pi$ from simulations (d-e). \textbf{b,} Zoomed-in region of (a) in the vicinity of the main CR overtone. \textbf{c-e} Numerical simulations of the near-field distribution of THz graphene plasmons (c)  and BMs above (d) and below (e) the cyclotron gap excited by a metal lead located in the contact with graphene sheet placed above the silicon gate at $\omega_c / 2 \pi =1~$THz for $n=0.6\times10^{12}$ cm$^{-2}$. Color scale  visualizes the real part of the vertical field component, $\Re E_\mathrm{z}(x,y)$. \textbf{f,} Calculated BM-assisted absorption of THz radiation as a function of $B$ normalized to $B_\mathrm{CR}$. Inset: model structure of graphene-metal interface which was used for calculations (See \SI{Supplementary Section {\color{blue} \ref{Magnetoabsorption_electrodynamics}}}). \textbf{g,} Photoresistance in the vicinity of the CR overtones measured at $f=0.69~$THz.   
	}
	\label{Fig3}
\end{figure*}

\textbf{Modeling THz magnetoabsorption.} Below we argue that the dominance of the CR overtones in the THz photoresponse in our device is a signature of the magnetoabsorption assisted by the excitation of the Bernstein modes (BMs) in graphene.  A qualitative picture of the effect is the following: Incident electromagnetic radiation is scattered by sharp highly conductive metal contacts to the structure, producing highly non-uniform near fields with large momenta ${\bf E}_{{\bf q}\omega}$. Each spatial harmonic is screened by electrons in graphene. For harmonics close to the dispersion of magnetoplasmon modes ${\bf q}_{\rm mp}(\omega)$, screening turns to anti-screening, or resonant field enhancement. This resonant near field becomes even stronger when the group velocity of plasmons tends to zero and their density of states becomes singular.
This occurs at the avoided crossing points of the CR overtones with the 'bare' plasmon dispersion $\omega_{\rm pl}({\bf q})$ obtained within the local approximation. The avoided crossings in the BM spectrum and the resonant absorption at CR harmonics originate from nonlocality of current-field response. Indeed, resonant electron transitions at multiples of $\omega_{\rm CR}$ are possible only in non-uniform electric fields.

The above qualitative considerations form a basis for a quantitative model of near-field absorption enhanced by graphene plasmons (See \SI{Supplementary Section \ref{Magnetoabsorption_electrodynamics}}). It enables us to calculate the absorbed power $P_{\rm near}$ due to scattered near fields, ${\bf E}_{{\bf q}\omega}$:

\begin{equation}
\label{eq1}
    P_{\rm near} = 2\int{\frac{d{\bf q}}{(2\pi)^2} \frac{\omega}{2 \pi q}|{\bf E}_{{\bf q}\omega}|^2 \Im \frac{1}{\varepsilon({\bf q},\omega)}}.
\end{equation}
In the above expression, ${\bf E}_{{\bf q}\omega} = F_{{\bf q}\omega} {\bf E}_0$ is the amplitude of ${\bf q}$-th spatial harmonic of electric near field, which is related to the incident field ${\bf E}_0$ via the diffraction amplitude $ F_{{\bf q}\omega}$, and $\varepsilon({\bf q},\omega)$ is the momentum- and frequency dependent dielectric function of graphene (\SI{Supplementary Section~\ref{Magnetoabsorption_microscopics}}). The so-called ``loss function'', $\Im \varepsilon^{-1}({\bf q},\omega)$, is responsible for the magnetoplasmon-assisted absorption. It peaks at the collective mode dispersion, as shown in Fig.~\ref{Fig3}a. 
The analysis of the collective mode contribution to the net power \eqref{eq1} shows that it is inversely proportional to the magnetoplasmon group velocity at that frequency, $v_\mathrm{gr}(\omega)$.
In an immediate vicinity of the anti-crossings, the group velocity approaches zero (See Fig.~\ref{Fig3}a, b). Here, the absorption is limited by the plasmon losses and the associated electron momentum relaxation time, $\tau_{\rm p}$, as $P_{\rm near} \propto \tau^{1/2}_{\rm p}$ (See \SI{Supplementary Section~\ref{Bernstein_modes}}).

The evaluation of the absorbed power for specific structure parameters requires the knowledge of two building blocks: the diffraction amplitude $F_{{\bf q}\omega}$ and the dielectric function $\varepsilon({\bf q},\omega)$. The former is evaluated by considering a plane wave scattering by a perfectly conducting rod that mimics a sharp contact (\SI{Supplementary Section~\ref{Magnetoabsorption_electrodynamics}}). The respective diffraction amplitude tends to zero at small $q$ due to the gate screening, reaches a maximum, and tends to zero at large $q$ due to the finite length and width of the contact. The dielectric function $\varepsilon({\bf q},\omega)$ is evaluated based on classical kinetic equation with weak particle-conserving collisions. The loss function calculated within this approach demonstrates well-defined Bernstein modes shown in Figs.~\ref{Fig3}a, b. 

To provide a more detailed picture of the BM resonance, in Figs.~\ref{Fig3}c-e we additionally show the real space distributions of the scattered electric field in different regimes. Both at zero magnetic field (panel c) and away from the BM plateau at $\omega \approx 2 \omega_{\rm CR}$ (panel d) contacts to the structure launch (magneto)plasmons with a well-defined real-space period of electric field corresponding to the wave vector $q_{\rm mp}(\omega)$~\cite{Alonso-Gonzlez-2014-spatial_conductivity_patterns}. Just below the maximum of the BM dispersion, there exist two magnetoplasmon modes with different $q$ at given radiation frequency (see Figs.~\ref{Fig3}a, b). This governs the short-period fringing of electric field threaded on long-period pattern in Fig.~\ref{Fig3}e. Exactly at the BM plateau, the two periods coalesce leading to the resonant enhancement of net absorbance.

\textbf{Discussion.} In Figs.~\ref{Fig3}f and g we compare the experimentally measured photoresistance and calculated magnetoabsorption (\SI{Supplementary section \ref{Bernstein_modes}}) in the vicinity of main CR overtones and find a good qualitative agreement. In particular,  both dependencies feature a strong asymmetry of the resonant peaks characterized by a rapid growth on the low-$B$ sides of CR overtones (similar to Figs.~\ref{Fig_1}d and ~\ref{Fig2}a). The reason for this asymmetry becomes clear when one realizes that the steep rise (at a short scale related to the scattering losses) corresponds to external frequency $\omega$ approaching the maximum of the BM dispersion from the gap side. The smooth drop at $B>B_{\rm CR}/2$ reflects the gradual increase of the group velocity of participating BMs for $\omega$ below the maximum (Fig.~\ref{Fig3}e). At that side, one returns to the conventional magnetoplasmon excitation regime, Fig.~\ref{Fig3}d, where the nonlocality, while present, does not play any essential role.

Remarkably, the asymmetry retains for the 3rd and 4th harmonics of the CR in both experimental and theoretical curves. This behaviour clearly distinguishes the BM resonances from the THz-induced magnetooscillations recently found in high-quality graphene - the oscillations there are readily described by a damped sinusoidal function in the limit of $B$-range applied in this work~\cite{VanyaRMP}. In this device, signatures of such sinusoidal magnetooscillations appear only at higher $f$ (\SI{Supplementary Section 3}). 

The key role in the excitation of Bernstein modes, similar to other collective modes in two dimensions, should be played by the inhomogeneity of electric fields due to the presence of sharp contacts.
To verify the idea, we have fabricated an additional device of comparable quality but contacted by metal leads outside the main channel (\SI{Supplementary Section 10}). 
We found strong photoresponse in the vicinity of the CR harmonics only for the invasive case, while for distant leads the photoresponse was dominated by THz-induced magnetooscillations~\cite{Monch2020}.

It is remarkable that the photovoltage and photoresistance at the main CR overtone can markedly exceed that at the CR itself. 
All the more, the fundamental resonance is observed only at low carrier densities or at high $f$ (\SI{Supplementary Sections 3 and \ref{Radiative_decay}}). The softening of the fundamental CR at large $n$ can be understood by taking into account the screening of incoming radiation by electrons in graphene, a dissipationless effect also referred to as radiative damping~\cite{Falko1989,Mikhailov2004,Muravev_APL_cyclotron_relaxation_time,SuperradiantKONO}. Calculations of the fundamental CR absorption based on transfer-matrix technique are shown in {Fig.~\ref{MainCR}} and capture this effect well. Absorption at CR overtones is not prone to screening, instead, it is enhanced by the near-field excitation of BMs. The observed decay of the second CR with increasing gate voltage is relatively weak. This can be attributed to reduced characteristic values of the magnetoplasmon wave vector $q^*$ and the respective decrease in the diffraction amplitude $F_{{\bf q}\omega}$.

It is instructive to briefly discuss other possible scenarios of the resonant photoresponce at CR overtones. In particular, similar resonances were predicted to emerge in the regime of viscous electron transport~\cite{Alekseev}. The aforementioned photoresponse asymmetry in our theory and experiments is, however, opposite to that obtained in Ref.~\onlinecite{Alekseev}, which predicts a smooth low-$B$ tail and a sharp high-$B$ tail of the magnetoabsorption peak around $B=B_\mathrm{CR}/2$.  Moreover, the hydrodynamic regime of electron transport in high density graphene sets in at $T>100~$K~\cite{BandurinFluidity} while the CR harmonics observed in this work disappear already at $T\sim 50~$K (Fig.~\ref{Fig2}c and \SI{Supplementary Section 4}). We have also addressed the effect of electron-electron (e-e) collisions on the BM-assisted magnetoabsorption and found that a fast collision rate results in reduction of the CR harmonics amplitudes (\SI{Supplementary Section 6}). Next, we note that although BMs were discussed in relation with the resonant photoresistance close to the 2nd CR harmonics observed in GaAs-based heterostructures~\cite{X2-peak-1}, the described scenario relied on the emergence of electron plasma instability mediated by the nonlinearity of the Euler equation and parametric resonance~\cite{Volkov2014}. The effect described here is linear in $P$ which excludes any instabilities-related mechanisms. The same observation also excludes multiphoton mechanisms which would inevitably lead to non-linear scaling with $P$~\cite{VanyaRMP}.

\textbf{Outlook.} Our results have several profound consequences for further research on nonlocal light-matter interaction at the nanoscale. First, the interaction between graphene plasmons and electron cyclotron motion via the formation of Bernstein modes and mediated by nonlocality effects can easily reach the strong coupling regime~\cite{Scalari_Ultrastrong_CR_THZ,Muravev_StrongCoupling}. Indeed, the BM gap due to resonant anticrossing  can be estimated as $\Delta \sim 10 (a_{\rm B}/R_{\rm c})^2 \omega_{\rm c}$, where $a_{\rm B}$ is the effective Bohr radius~\cite{Volkov2014}. Taking $f=1.63$ THz, we find $\Delta/\omega_{\rm c}\sim 0.17$ which is well beyond the collision-induced broadening of the resonance. In this regime, electron relaxation from higher Landau levels may demonstrate coherent quantum Rabi oscillations that can be potentially detected in time-resolved measurements~\cite{Belyanin_Time-resolved_CR}. Second, that ultra-slow collective excitations associated with the Bernstein modes close to the CR harmonics can be sensitive to many-body effects~\cite{LevitovFeigelman}. Our approach thus paves the way to access these effects in simple far-field experiments. Last but not least, our observations revisit the role of nonlocal conductivity in light-matter interaction, that was previously believed to hamper field compression and slowing of light~\cite{Ciraci-2012-Limits_of_Plasmonic_Enhancement}. Our study refutes this perspective by revealing highly-confined ultra-slow plasmon modes enabled by nonlocality.




\vspace{1em}

\noindent\rule{6cm}{0.4pt}

*Correspondence to: bandurin@mit.edu \\ svintcov.da@mipt.ru, sdg51t@gmail.com

\section*{Methods}

\subsection*{Device fabrication}

To fabricate our devices we first encapsulated graphene between relatively thick hBN crystals using hot-release technique \cite{Purdie2018}. The stack was transferred either on top of a predefined back gate electrode made of graphite or onto Si$^{++}$/SiO$_\mathrm{2}$ wafer. The resulting van der Waals heterostructure was patterned using electron beam lithography to define contact regions which followed by reactive ion etching (RIE) to selectively remove the areas unprotected by a lithographic mask, resulting in trenches for depositing electrical leads (3nm of chromium, 60 nm of gold). For one of our devices, we used rectangular shape graphene flake which formed the device channel whereas for another device, the graphene channel was defined by a third round of e-beam lithography, followed by RIE using Poly(methyl methacrylate) and gold top gate as the etching mask.

\subsection*{Measurement technique}

The sample was mounted within a variable temperature inset of an optical cryostat equipped with $z$-cut crystal quartz windows, which were covered by a black polyethylene (PE) film. While the latter is transparent for THz radiation the PE film functions as a filter in both visible and infrared ranges preventing uncontrollable excitation of graphene by room light. The homogeneous magnetic field, $B$, was generated by superconductive coils in the range $\pm$7~T. Standard four-terminal magnetotransport measurements were performed using $ac$ currents between 1 and 5~µA providing information about the sample's transport characteristics and quality.

Our high-quality graphene devices were investigated under several frequencies, namely $f = 0.69$, 1.63 and 2.54~THz, with corresponding photon energies, $\hbar\omega$ = 2.9, 6.7 and 10.5~meV, generated by a continuous wave ($cw$) optically pumped molecular laser using CH$_2$O$_2$, CH$_2$F$_2$ and CH$_3$OH, respectively, as active media. The laser beam was recorded by a pyroelectric camera,~\cite{Ziemann2000} which showed a nearly Gaussian profile with FWHM spot diameters $d_\mathrm{s}$ = 2.58, 1.75 and 1.56~mm at the sample's position. The incident average powers, $P$, lied in the range from 15 to 80~mW providing intensities, $I_0 = P/A_\mathrm{s}$, up to 3.32~W/cm$^{2}$.~\cite{Danilov2009, Olbrich2013} The state of the initially linear polarized radiation was controlled in angle and ellipticity by rotation of wave plates made of $x$-cut crystal quartz. All measurements were performed in Faraday geometry, where both the magnetic field, $B$, and the incoming THz radiation were perpendicular to the graphene plane.

In this work the photoresponse induced by the illumination with THz radiation was obtained by either photovoltage or photoresistance measurements. In the photovoltage regime the induced signal was measured as a voltage drop $V_\mathrm{ph}$ between a pair of contacts without applying external bias to the device. The incident radiation was modulated by an optical chopper to provide a phase sensitive detection with standard lock-in technique.

In order to obtain a photoresistance signal, i.e. the change of $dc$ conductivity induced by THz radiation, an $ac$ current, $I_\mathrm{ac}$, between 5 and 50~µA was applied to the device with a modulation frequency lying within the range $f_\mathrm{ac}$ = 5 - 12~Hz, while the sample was exposed to laser radiation simultaneously modulated at a significantly larger frequency of $f_\mathrm{chop} = 140$~Hz (see, e.g., Ref.~\onlinecite{Kozlov2011}). Such double-modulation measurement technique is realized via two lock-in amplifiers connected in series. Here, the first lock-in is phase-locked to $f_\mathrm{chop}$ extracting the total signal induced by THz radiation. The amplitude of the total photosignal, oscillating at $f_\mathrm{chop}$, includes a constant component proportional to the photovoltage and an alternating part modulated with $f_\mathrm{ac}$. This total signal feeds the second amplifier which is locked to $f_\mathrm{ac}$ thus generating a voltage proportional to the photoresistance.


\subsection*{THz absorption modelling}
The absorption cross-section was evaluated using the Joule's law relating the absorbed power, the real part of the dynamic 2D conductivity, and local electric field.  The wave-vector dependent dynamic conductivity was obtained by solving the classical kinetic equation for 2D electron gas in magnetic field in the presence of weak disorder~\cite{Chaplik_1985_CR_overtones}. The resulting expression for $\sigma({\bf q},\omega)$ is a sum of terms resonant at various CR overtones, with $q$-dependent amplitudes. Electron-impurity collisions were included in a particle-conserving way; for e-e collisions we used a model particle- and momentum-conserving collision integral~\cite{Svintsov_PRB_Crossover}. Electric field acting on electrons in graphene is a sum of an incident plane wave and evanescent wave scattered by the keen contact. The amplitude and shape of the evanescent wave were found by solving the scattering problem for conducting rod~\cite{Hallen_JAP_LongAntenna}. Both plane and evanescent wave amplitudes are modified due to screening by electrons in graphene. As a result, each ${\bf q}$-th spatial harmonic of the field is divided by 2D dielectric function $\varepsilon({\bf q}, \omega) = 1 + 2 \pi i \sigma({\bf q},\omega)\sqrt{q^2 - k_0^2}/\omega $, where $k_0 = \omega/c$ is the wave number of incident wave. The effective dielectric function modified by the presence of encapsulating dielectric and gate is given in the \SI{Supplementary Section 6}.

\section*{Acknowledgments}
Regensburg team acknowledges the support of the Deutsche Forschungsgemeinschaft (DFG, German Research Foundation) – Project-ID 314695032 – SFB 1277 (Subproject A04). I.A.D. acknowledges the DFG support via grant DM 1/5-1. Work at MIT was partly supported through AFOSR Grant FA9550-16-1-0382, through the NSF QII-TAQS program (Grant 1936263), and the Gordon and Betty Moore Foundation EPiQS Initiative through Grant GBMF9643 to P.J.H. D.A.B. acknowledges the support from MIT Pappalardo Fellowship. I.Y.P acknowledges support from the MIT undergraduate research opportunities program and the Johnson \& Johnson research scholars program. Support from the Materials Engineering and Processing program of the National Science Foundation, award number CMMI 1538127 for hBN crystal growth is also greatly appreciated. The work of D.S. was supported by the Foundation for Advancement of Theoretical Physics ''Basis'', grant \# 20-1-3-43-1. We thank Leonid Levitov, Clement Collignon and Alexey Berdyugin for valuable discussions and Alexander A. Zibrov for his sharing experience in the fabrication of ultra-high quality graphene devices. 

\section*{Data availability}
All data supporting this study and its findings are available within the article and its Supplementary Information or from the corresponding authors upon reasonable request.

\section*{Author contributions}
D.A.B. and S.D.G conceived and designed the project. E.M., D.A.B. and K. L. performed the transport and photoresponse measurements. D.A.B. and I.Y.P. fabricated the devices. D.A.B., E.M., I.A.D., D.S. analyzed the experimental data with the help from P.J.H. and S.D.G.. K.K. and D.S. developed theoretical model and performed magnetoabsorption calculations. S.L., J.H.E. provided high-quality hBN crystals. D.A.B., I.A.D., E.M., and D.S. wrote the manuscript with input from all co-authors. P.J.H. and S.D.G. supervised the project. All authors contributed to discussions. 
\newline

\section*{Competing interests}
The authors declare no competing interests.




\clearpage
\newpage
\setcounter{figure}{0}
\setcounter{section}{0}
\setcounter{equation}{0}
\renewcommand{\thesection}{}
\renewcommand{\thesubsection}{\arabic{subsection}}
\renewcommand{\thesubsubsection}{\arabic{subsection}.\arabic{subsubsection}}
\renewcommand{\theequation} {S\arabic{equation}}
\renewcommand{\thefigure} {S\arabic{figure}}
\renewcommand{\thetable} {S\arabic{table}}

\begin{widetext}

\section*{Supplementary Information}


\subsection{\textbf{
Device characterization}}
Prior to the photoresponse measurements, we characterized the transport properties of our device by measuring its  resistivity $\rho_\mathrm{xx}$ as a function of gate voltage $V_\mathrm{g}$ at $B=0$ (Fig.~\ref{Fig_1}e). Such measurements allowed us to determine three important interrelated characteristics - the mobility, $\mu$, the transport scattering time, $\tau_\mathrm{p}=\mu m/e$, and the mean free path, $l_\mathrm{mfp}=v_\mathrm{F}\tau_\mathrm{p}$, using the standard Drude relation $\rho_\mathrm{xx}=1/e n \mu$.
The obtained characteristics reveal high-mobility electron transport, with $\mu$ exceeding $10^6$ cm$^2$/Vs at low densities $n$ (Fig.~\ref{FigS1}a), and yield $\tau_\mathrm{p}\simeq 8$~ps that is approximately independent of $n$ (Fig.~\ref{FigS1}b). The latter value gives $l_\mathrm{mfp}\simeq 8~\mu$m accurately matching the width of our device (Fig.~\ref{Fig_1}e). We conclude that at $B=0$ the electron transport is mainly determined by scattering on edges of graphene. To test this assumption further, we applied a perpendicular magnetic field $B$ and found a strong negative magnetoresistance shown in Fig.~\ref{FigS1}c. Such magnetoresistance is routinely explained~\cite{MR_graphene,BEENAKKER} in terms of skipping orbits that are forming in magnetic field. With increasing $B$, the cyclotron orbits eventually become smaller than the width of device eventually leading to transition to a diffusive transport regime where the transport is not affected by the backscattering at the edges.

\begin{figure*}[ht!]
	\centering\includegraphics[width=0.8\linewidth]{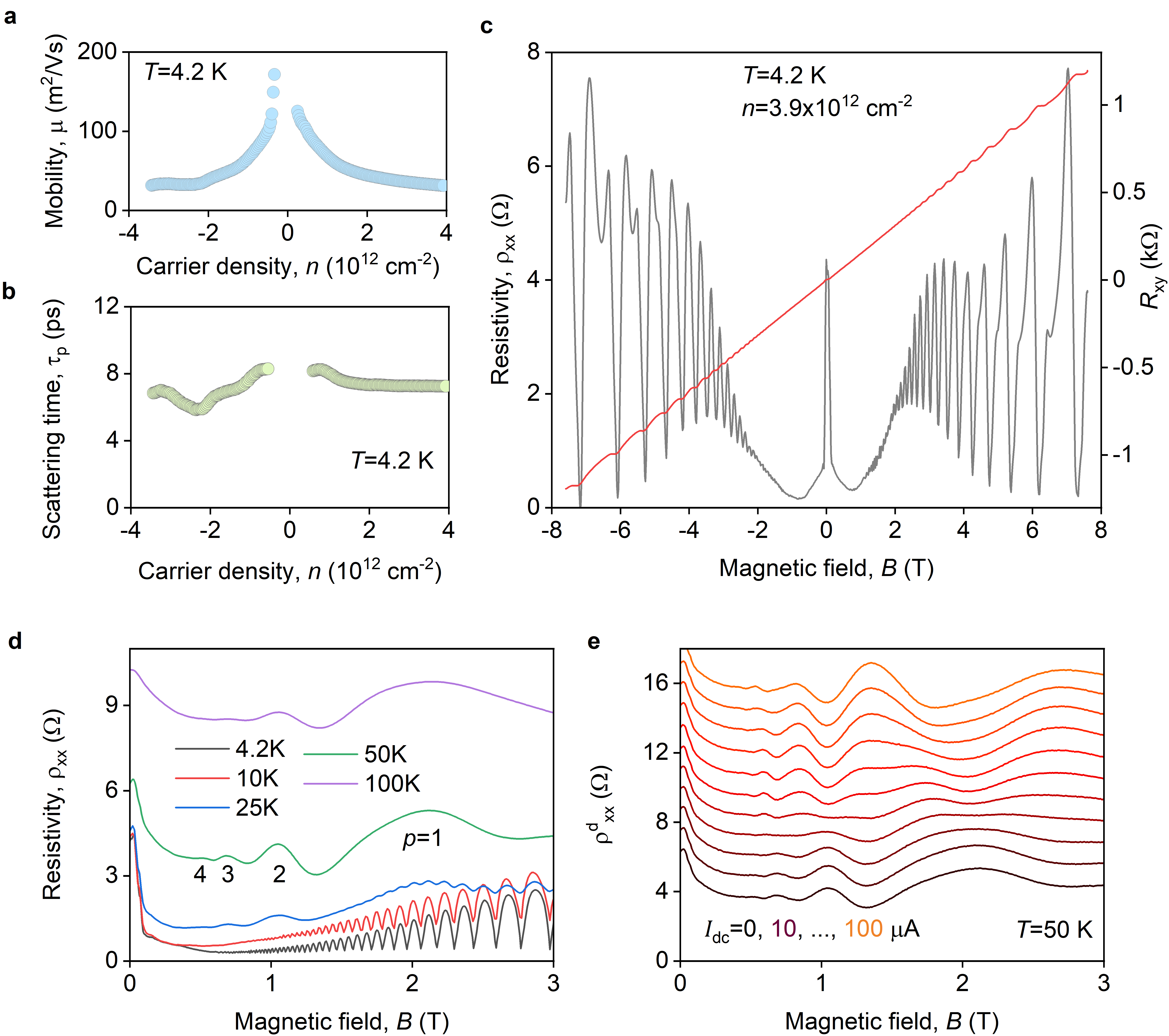}
	\caption{\textbf{Transport properties.} \textbf{a-b,} Mobility $\mu$ and transport scattering time $\tau_{\rm p}$ as a function of carrier density $n$ at given $T$. \textbf{c,} \EM{Panel (c) dividing factor of 2 is missing for $\rho_{xx}$, $s=0.5$} 
	Linear longitudinal resistivity $\rho_\mathrm{xx}$ and Hall resistance $R_\mathrm{xy}$ versus magnetic field $B$ measured at given $n$ and $T$.  \textbf{d,}  $\rho_\mathrm{xx}$ at given temperatures demonstrating the temperature evolution of PIRO and SdHO. 
	$p=1$, 2, \ldots mark the maxima of PIRO.\cite{zudovPIRO2001,raichevPIRO2009,RoshMP,VanyaRMP} Individual traces are not vertically shifted. Together with PIRO, the visible overall growth of resistivity with $T$ reflects the increasing role of thermal acoustic phonons. \textbf{e,} Differential resistivity $\rho^\text{(d)}_\text{xx}\equiv s(\text{d}V_\text{x}/\text{d}I_\text{x})_{I_\text{x}=I_\text{dc}}$ versus $B$ for different values of injected direct current $I_\text{dc}$ (here \sout{$s=0.5$} \EM{for panel (d) and (e) the geometric aspect ratio is $s=1$; change ?$s \rightarrow a$?, since $s$ is already the speed of sound, see main text PIRO} 
	is the geometrical aspect ratio). The traces, vertically shifted for clarity, demonstrate evolution of PIRO under strong dc bias similar to that reported earlier\cite{ZhangPHIRO2008,DmitrievPHIRO2010} in high-mobility 2DES based on GaAs/GaAlAs heterostructures.}
	\label{FigS1}
\end{figure*}

A further confirmation of transition to a diffusive transport at $B$ exceeding $0.1$~T are the phonon-induced resistance oscillations (PIRO) which were observed at elevated $T>10$~K, see Fig.~\ref{FigS1}d. These oscillations, previously mostly studied in high-mobility 2DES with parabolic dispersion\cite{zudovPIRO2001,raichevPIRO2009,VanyaRMP}, were recently detected in high-quality monolayer graphene samples.\cite{RoshMP} The previous studies established that PIRO reflect the commensurability of the cyclotron energy $\hbar\omega_\text{c}$ with the acoustic phonon energy $E_{2 p_\text{F}}=2 p_\text{F} s$ required to back-scatter an electron at the Fermi surface (here $p_\text{F}$ is the Fermi momentum and $s$ the speed of sound). Due to contribution of such scattering events at magnetic fields above the onset of Landau quantization and temperatures sufficient to provide thermal phonons of energy $E_{2 p_\text{F}}$, the resistance is resonantly enhanced at $p\equiv E_{2 p_\text{F}}/\hbar\omega_\text{c}=1,2,\ldots$, see Fig.~\ref{FigS1}d. The resulting PIRO are $1/B$-magnetooscillations which feature a non-monotonic temperature dependence and survive at high $T$ exceeding 100~K, in contrast to conventional Shubnikov-de Haas oscillations (SdHO) which, at relevant $B$, get strongly thermally suppressed already at 25 K, see Fig.~\ref{FigS1}d. 

Figure \ref{FigS1}e demonstrates strong modifications of PIRO induced by application of a moderately strong direct current $I_\text{dc}$ along the Hall bar. One observes that, with increasing $I_\text{dc}$, the successive maxima $p=1,2,\ldots$ PIRO flatten out and subsequently turn into minima, and vice versa. Such behavior was detected earlier in 2DES with parabolic dispersion\cite{ZhangPHIRO2008} and was well reproduced by theory taking into account the tilt of Landau levels in  the Hall field induced by the applied direct current.\cite{DmitrievPHIRO2010}  Indeed, the phonon-assisted backscattering leading to PIRO corresponds to a well-defined spatial shift of the cyclotron orbit by a distance equal to the cyclotron diameter; in the presence of the dc Hall field, such spatial shifts change the electrostatic energy of participating electrons and, therefore, modify the resonance conditions for the inter-Landau-level phonon-assisted backscattering.\cite{VanyaRMP} These results unequivocally demonstrate the bulk diffusive nature of the dc transport in the magnetic field range relevant for phenomena discussed in the manuscript, and additionally attest an excellent quality of our devices. A detailed analysis of the above and related phenomena in graphene will be presented elsewhere.




\clearpage
\newpage

\subsection{\textbf{
Further examples of the THz-driven magnetoresistance}}

The anomaly in the THz-driven magnetoresistance in the vicinity of the main overtone of CR was found universal and revealed itself for different $n$. Figure~\ref{FigS2MR} shows further examples of $\Delta R$ dependencies on $B$ in the dark (black) and in the presence of $0.69~$THz radiation (red) measured at two representative values of $n$. Pronounced resistance spikes were observed close to the position of the CR main overtone (arrows) with no features close to position of the principal CR. 

\begin{figure*}[ht!]
	\centering\includegraphics[width=0.8\linewidth]{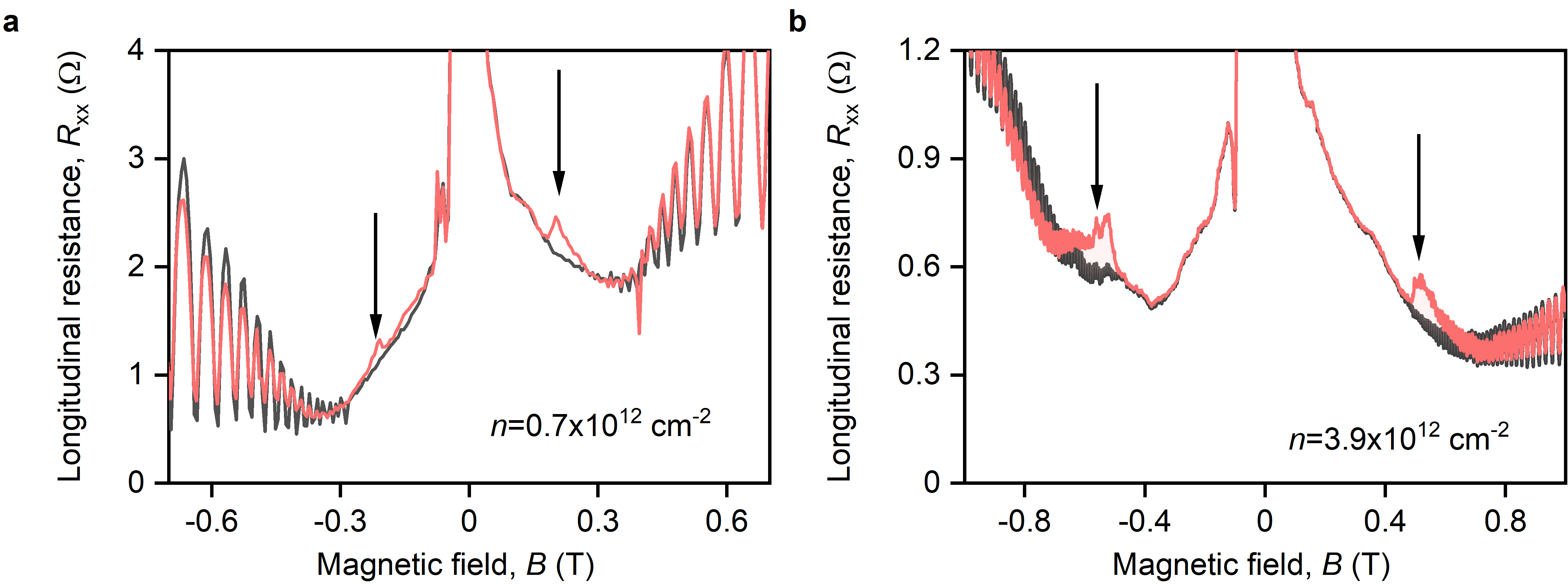}
	\caption{\textbf{Further examples of THz-driven magnetoresistance.} \textbf{a-b,} Longitudinal resistance $R_\mathrm{xx}$ versus magnetic field $B$ at different $n$: $0.7\times10^{12}~$cm$^{-2}$ (a) and $3.9\times10^{12}~$cm$^{-2}$ (b) measured in the dark (black) and under 0.69~THz radiation (red). $T=4.2~$K. Arrows indicate the positions $B=B_\text{CR}/2$ of the CR main overtone. }
	\label{FigS2MR}
\end{figure*}

\clearpage
\newpage

\subsection{\textbf{
Carrier density and frequency dependence}}

To fully characterize the observed anomalous behaviour, we explored the whole parameter space and studied the photoresponse dependencies on the carrier density $n$ and frequency $f$ of incident radiation using the double modulation technique described in Methods. The results are reported in Fig.~\ref{FigS3nANDf_dep} that shows $\Delta R$ as a function of $B$ for varying $n$ and three different $f=0.69$ (a), $1.63$ (b) and 2.54~THz (c). Increase of the $\Delta R$ close to the CR main overtone was observed at all $f$ with somewhat stronger effect at the smallest $f=0.69~$THz (Fig.~\ref{FigS3nANDf_dep}a). At the position of the principal CR, we observed photoresistance increase only at small $n$ close to the CNP or at high $f$. We attribute such behavior to the screening effect (\SI{Supplementary Section 8}). Last but not least, at the highest $f=2.54~$THz we detected THz-induced magnetooscillations~\cite{VanyaRMP}. The latter has been recently discovered in graphene in the photovoltage measurements yet until now no signature of TIMO were detected in the resistance studies. Figure~\ref{FigS3nANDf_dep} reveals this effect in our device's resistance making TIMO an ubiquitous phenomenon in high-quality graphene exposed to THz radiation.

\begin{figure*}[ht!]
	\centering\includegraphics[width=0.9\linewidth]{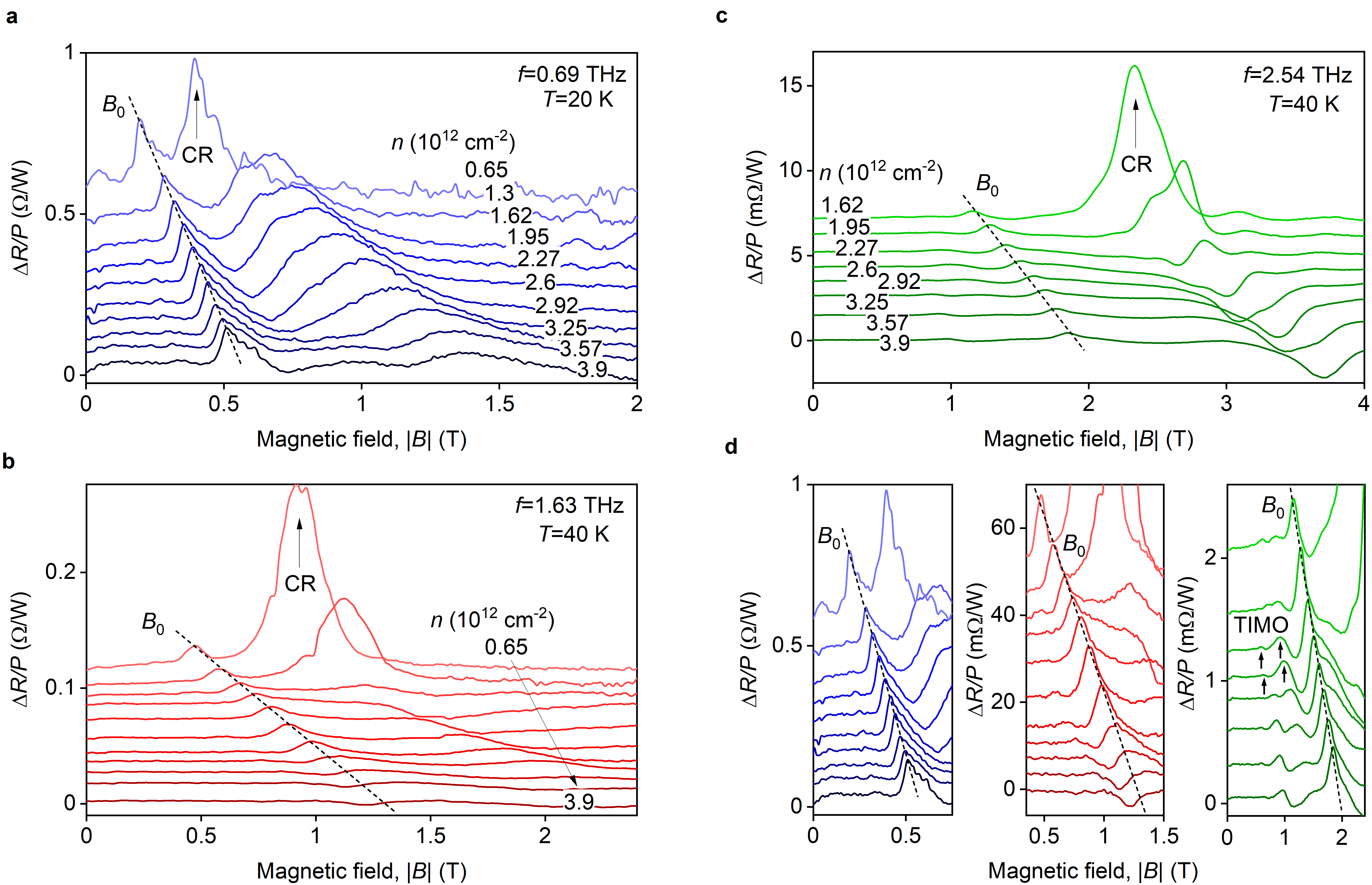}
	\caption{\textbf{$n-$ and $f-$ dependencies of the photoresistance.} \textbf{a-c,} Photoresistance $\Delta R$ as a function of magnetic field $B$ for varying $n$ at $f=0.69~$THz (a), $f=1.63~$THz (b) and $f=2.54~$THz (c) measured at given $T$. $B_0$ and CR label the photoresponse peak in the vicinity of the CR main overtone and that emerging at the position of the principal cyclotron resonance. Note that CR peak drops drastically with increasing $n$ whereas $B_\mathrm{0}$ experiences much slower decay with practically no apparent dependence at $f=0.69~$THz. \textbf{d,} Zoomed-in photoresponse from (a-c) showing  the details of the $\Delta R(B)$ dependence in the vicinity of $B_0$. At the highest $f=2.54~$THz in our experiments, $\Delta R$ features additional effect: TIMO. The latter until now has not been observed in graphene devices in the photoresistance measurements and revealed itself only in the photovoltage studies~\cite{Monch2020}. }
	\label{FigS3nANDf_dep}
\end{figure*}

\clearpage
\newpage

\subsection{\textbf{
Temperature and power dependence}}
Figures~\ref{FigS4-1}a-b show the full data set for the $\Delta R(B)$ dependence on temperature, $T$, measured under illumination of the graphene device with $f=0.69~$THz radiation using the double modulation technique described in Methods. Sharp asymmetric peaks in the $\Delta R(B)$ data, pronounced the most at the liquid helium $T$, are suppressed upon raising $T$ and completely vanish at $T=50~$K. The zoomed-in data plotted in Fig.~\ref{FigS4-1}b also reveals similar behavior for the second and third overtones (labeled as $B_\mathrm{CR}/3$ and $B_\mathrm{CR}/4$,  respectively) with somewhat faster drop. At $T>30~K$, the $\Delta R(B)$ dependencies feature additional oscillating pattern for $B>B_\mathrm{CR}$ of unknown origin. We attribute them to the manifestation of the phonon-induced magnetooscillations recently discovered in graphene devices of comparable quality~\cite{RoshMP}.  

\begin{figure*}[ht!]
	\centering\includegraphics[width=0.6\linewidth]{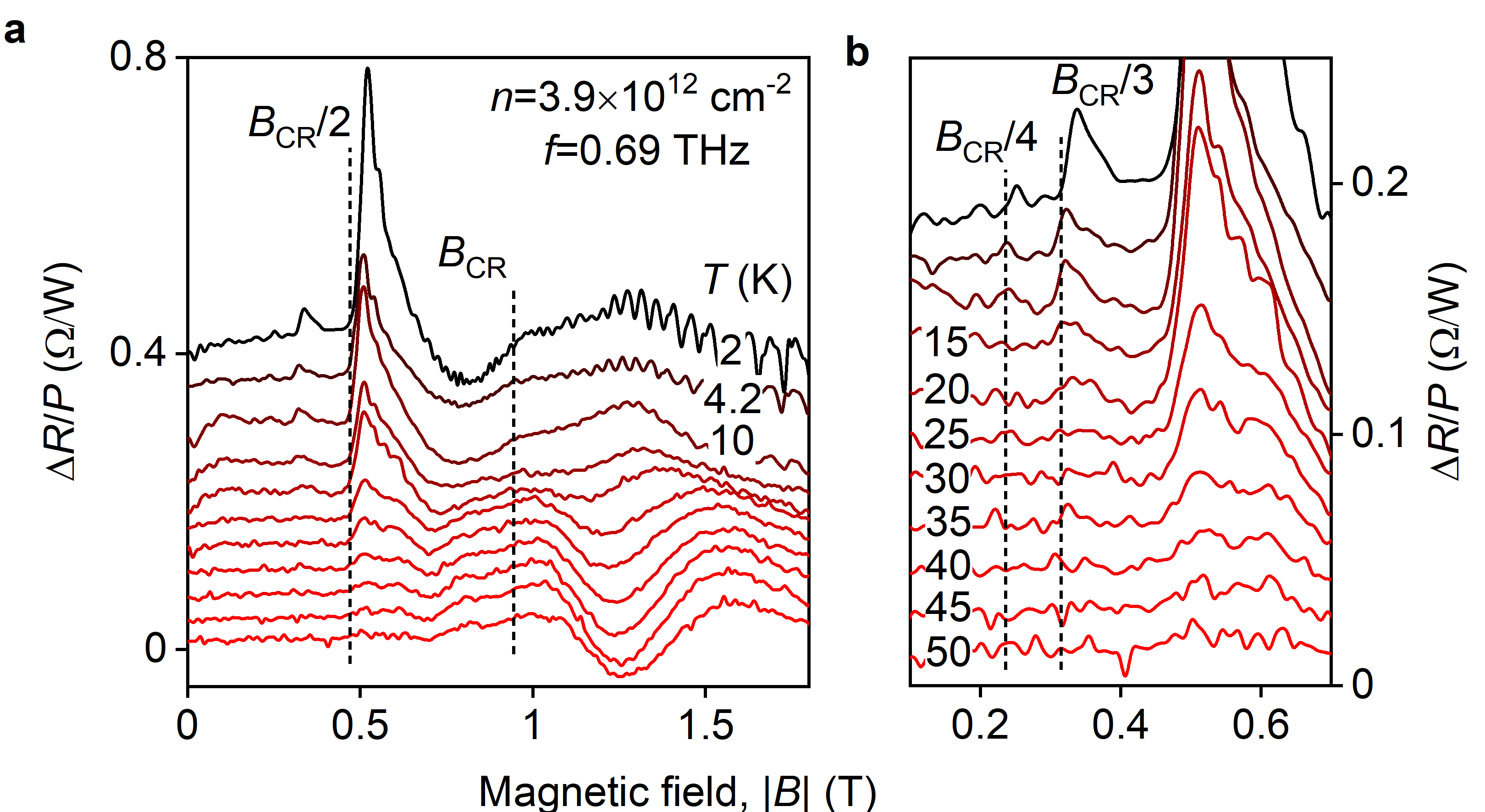}
	\caption{\textbf{Temperature dependence of the photoresistance in the vicinity of the CR overtones.} \textbf{a,} $\Delta R$ as a function of magnetic field $B$ at different $T$. \textbf{b,} Zoomed-in $\Delta R(B)$ dependencies showing the thermal damping of the second and third CR overtones labeled as $B_\mathrm{CR}/3$ and $B_\mathrm{CR}/4$. }
	\label{FigS4-1}
\end{figure*}

We also studied the dependence of the anomalous photoresponse of our graphene device in the vicinity of the CR main overtone on the intensity, $I_\mathrm{0}$, of the incident THz radiation. Figure~\ref{FigS4-2} shows the result of such study in the form of $\Delta R(B)$ traces measured at different $I_\mathrm{0}$. The analysis shows that the height of the $\Delta R$ peak close to the CR main overtone is proportional to $I_\mathrm{0}$ at small intensities (see Fig. 2d of the main text). This indicates that the origin of the observed effect is not related to the instability-related mechanisms previously called to address photoresistance spike in high-mobility GaAs-based 2DES~\cite{Volkov2014}. 

\begin{figure*}[ht!]
	\centering\includegraphics[width=0.5\linewidth]{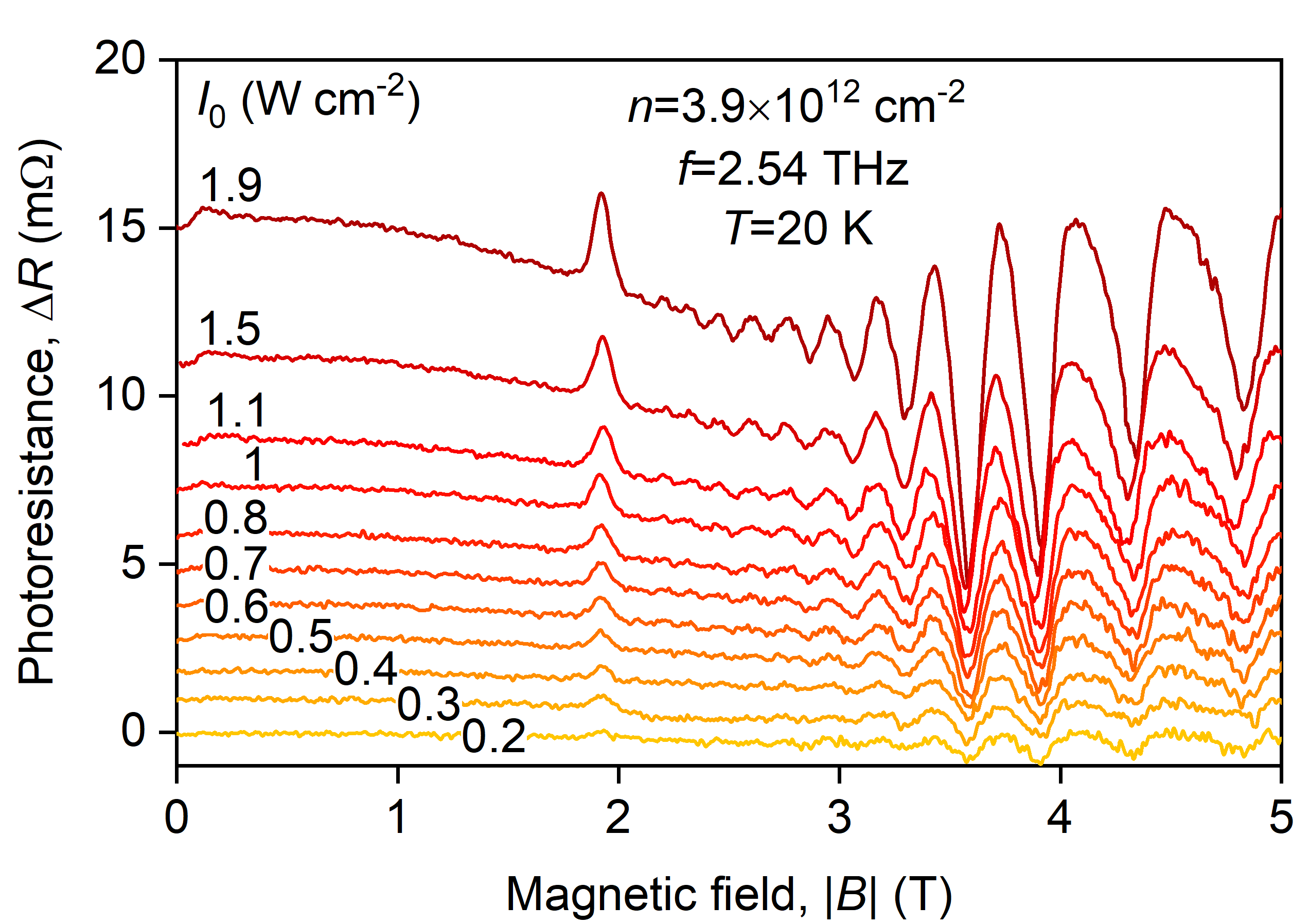}
	\caption{\textbf{ 
	 Intensity dependence of the photoresistance. 
	}
	Photoresistance $\Delta R$ as a function of magnetic field $B$ measured in response to $2.54~$TH radiation at given $n$ and $T$ and different intensities $I_0$.  
	}
	\label{FigS4-2}
\end{figure*}

\clearpage
\newpage

\subsection{\textbf{
Polarization dependence }}

The absorption of the electromagnetic radiation via the CR is in general sensitive to the polarization of the incident wave and the direction of magnetic field with respect to the 2DES plane. The magnetic field direction defines whether electrons experience clockwise or anticlockwise cyclotron motion and, thus, magnetoabsorption with only a particular polarization (left or right) should be active. 

We have tested the above scenario in our graphene device and performed the photovoltage measurements upon varying the beam polarization. For this measurements we have chosen the highest $f$ available in our experimental setup and fixed other parameters such that both the fundamental CR and its main overtone are observed simultaneously. Figure~\ref{FigS5} shows $\Delta V_\mathrm{ph}$ as a function of $B$ measured in response to $2.54~$THz linearly polarized radiation. Independently on the direction of $B$, $|\Delta V_\mathrm{ph}|$ peaks at the principal CR and its main overtone with two curves practically overlapping each other. The situation changes drastically for the circularly polarized radiation:  $\Delta V_\mathrm{ph}(B)$ features a clear asymmetry with respect to the $B-$field direction for different polarization helicities.  More precisely, the left-hand (LCP) and right-hand (RCP) circularly polarized (LCP) beams cause a strong $\Delta V_\mathrm{ph}$ dip for the negative and positive $B$ polarities, respectively.

This standard behavior is modified, however,  in the vicinity of the CR main overtone: here, $V_\mathrm{ph}(B)$ dependencies are both insensitive to the polarization angle of the linearly-polarized beam and unaffected by the radiation chirality (Insets of Figs.~\ref{FigS5}a-b). This suggests that the magnetoabsorption in the vicinity of the CR main overtone is, in contrast to conventional CR, indeed, a near-field effect occurring close to the sharp contacts  whose geometry determines the local polarization. Our theoretical model presented in Supplementary Section 8 captures this effect well. 

\begin{figure*}[ht!]
	\centering\includegraphics[width=0.8\linewidth]{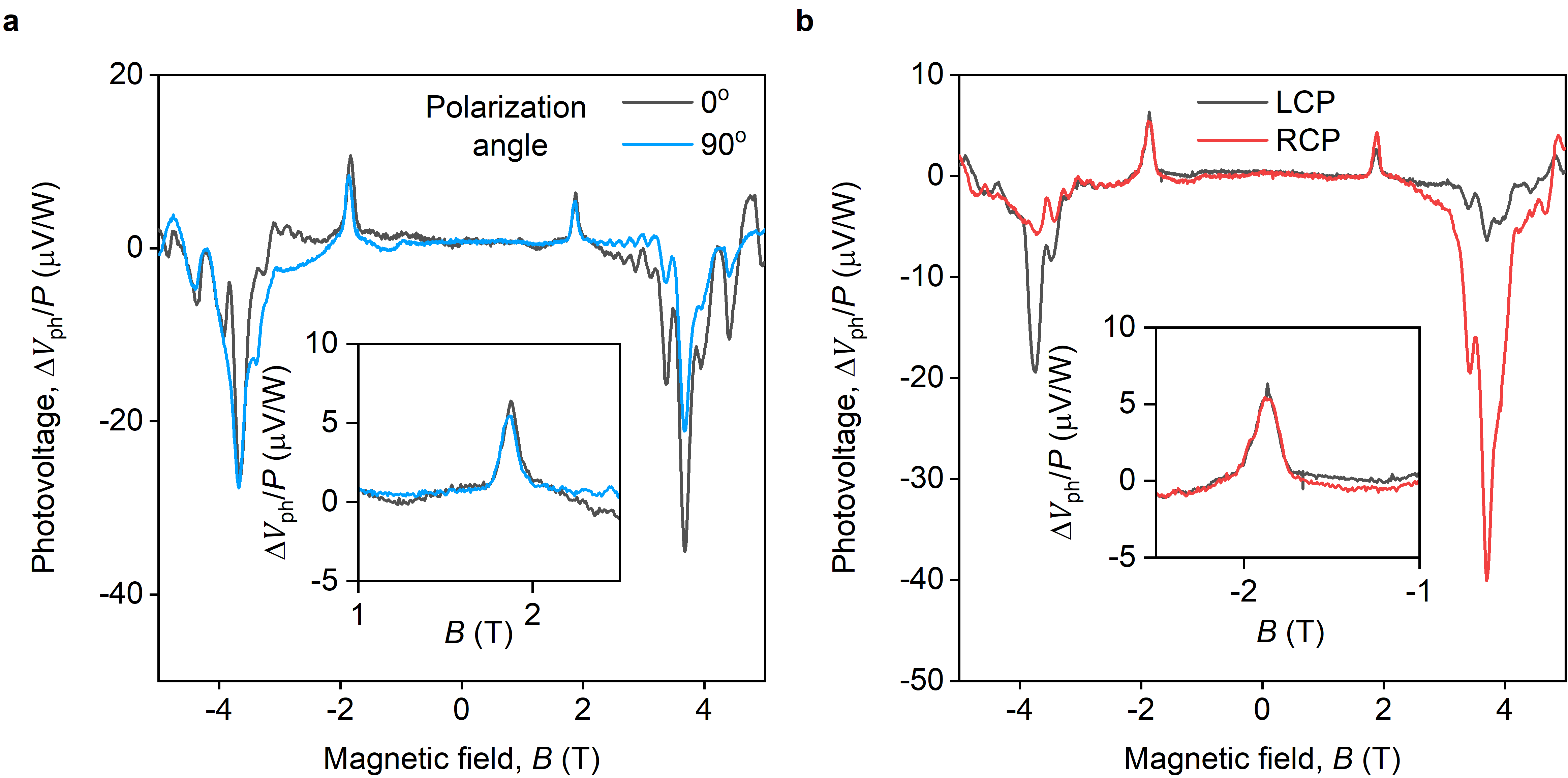}
	\caption{\textbf{Polarization dependence of the principal CR and its main overtone.} \textbf{a,} Photovoltage $V_\mathrm{ph}$ as a function of magnetic field $B$ measured in our device in response to incident linearly-polarized $f=2.54~$THz radiation for two orthogonal polarization angles measured with respect to the Hall bar long axis. Inset shows the response in the vicinity of the CR main overtone. \textbf{b,} $V_\mathrm{ph}$ versus $B$ for the same $f$ and $n$ as in (a) measured in response to the left and right circularly polarized radiation (LCP and RCP respectively).  Inset: zoomed-in $V_\mathrm{ph}$ close to the CR main overtone.}
	\label{FigS5}
\end{figure*}

\clearpage
\newpage

\subsection{\textbf{
Magnetoabsorption and electrodynamics of near fields}}
\label{Magnetoabsorption_electrodynamics}

\subsubsection{Magnetoabsorption}
To show the plasmonic nature of resonances at multiples of cyclotron frequency, we study the absorption of a highly inhomogeneous field by graphene. In our model, we consider the absorption of an inhomogeneous field with Fourier harmonics $\mathbf{E}(\mathbf{q} \omega)$ by an infinite two-dimensional electron system. The assumption of extended 2DES is applicable of its size is well above the plasmon wavelength. Provided that the spatial Fourier harmonics of the inhomogeneous field and the current density are known, the absorbed power in a two-dimensional sample is given by Joule's law
	\begin{gather}
	\label{Q}
	    P = 2 \int \Re{[\mathbf{E} \mathbf{j}^*]}d \mathbf{q}\,.
	\end{gather}

Further, we take into account the Ohm's law  $\mathbf{j} = \hat{\sigma} \mathbf{E}$ with conductivity tensor \begin{equation}
\hat{\sigma} =
	  \begin{pmatrix}
                    \sigma_{xx} & -\sigma_{xy} \\
                    \sigma_{xy} & \sigma_{xx}
    \end{pmatrix}    ,
\end{equation}
and obtain an expression for the absorbed power in the form
    \begin{gather}
        \label{Absorp}
        P = 2 \int \dfrac{d \mathbf{q}}{(2 \pi)^2} \left(\sigma'_{xx} |\mathbf{E}(\mathbf{q}, \omega)|^2 + \sigma''_{xy} \Im{[E^*_{ x} (\mathbf{q}, \omega)E_{y}({\mathbf{q}, \omega})]}\right),
    \end{gather}
where prime and double prime stand for real and imaginary parts. The second term in Eq.~\ref{Absorp} is responsible for the polarization sensitivity of cyclotron resonance. We shall further prove its importance for the main CR excited via far-field harmonics. For the CR overtones excited by the near-field harmonics, the polarization of radiation is very close to the linear one, independent of the state of the incident light. Indeed, the component of the field normal to the contact surface is enhanced via lightning-rod effect, while the tangential component is screened via skin-effect in metal. A more rigorous justification of this statement can be obtained via exact solution for diffraction at a semi-infinite metal sheet (see subsection 'Diffracted wave polarization'). As a result, the second term in Eq.~(\ref{Absorp}) can be safely neglected for the CR overtones.

\subsubsection{Diffracted field}
One of the main blocks required for calculation of absorption is the wave-vector dependent (inhomogeneous) field $\mathbf{E}_d(\mathbf{q},\omega)$ arising due to the diffraction of the incident wave at the contacts to the structure. To find the diffracted field, we model the true contact as a thin perfectly conducting rod of length $2L$ and radius $a$ ($a \ll L$) (Fig.~\ref{Fig_rod}).
\begin{figure}[h!]
\begin{minipage}[h]{1\linewidth}
\center{\includegraphics[width=0.85\linewidth]{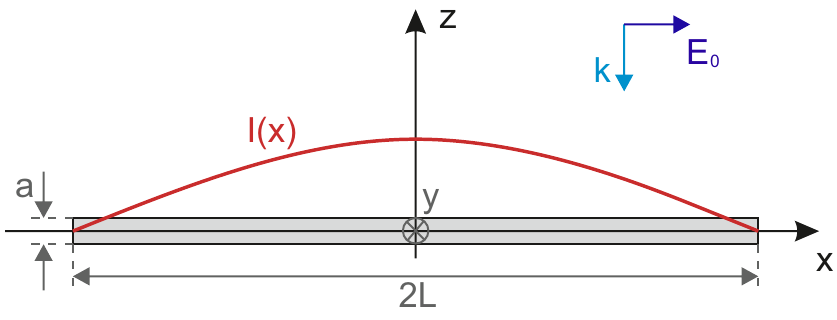} }
\caption{Schematic of the model contact, a perfectly conducting rod of length $2L$. Red solid line shows the distribution of electric current along the rod.}
\label{Fig_rod}
\end{minipage}
\end{figure}
The rod is located in $xy$-plane of the 2DES, the latter is assumed to be infinite. An incident plane electromagnetic wave is propagating against the $z$-axis. The field of the incident wave $\mathbf{E}_0$ induces high-frequency current $I(x)$ distributed along  the rod. The distribution of current $I(x)$ can be found from Pocklinton integral equation with boundary conditions $I(\pm L)=0$~\cite{pocklington1897electrical}. The latter states that electric field induced by the currents should exactly compensate the incident field in the case of perfectly conducting wire. Further on, for thin wires $a \ll L$ it is possible to use local capacitance approximation for the vector-potential induced by this current:
\begin{gather}
\label{A}
A_{ind,x}(x) = \dfrac{1}{4 \pi c} \int_{-L}^L I(x') \dfrac{e^{-i k \sqrt{(x-x')^2+a^2}}}{\sqrt{(x-x')^2+a^2}} dx' \approx   I(x) \dfrac{2 \ln{\frac{2L}{a}}}{4 \pi c}.
\end{gather}
\\
Above, $\kappa \equiv 2 \ln{\frac{2L}{a}}$ is the effective local capacitance of the wire. Relating the induced electric field with electric potential in the Lorentz gauge, we find a simple differential equation for $A_{ind,x}(x)$:
\begin{gather}
\label{eq}
E_{ind,x} = \dfrac{i}{k} [ A_{ind,x}''(x) + k^2 A_{ind,x}(x) ] = - E_{0}.
\end{gather}
Equation (\ref{eq}) is readily solved with zero-current boundary conditions $I(\pm L) = 0$ and leads to the following distribution of current:
\begin{gather}
\label{I}
    I(x) = i \dfrac{4 \pi E_0 c^2}{\kappa \omega} \left( 1 - \dfrac{\cos{k x}}{\cos{k L}} \right).
\end{gather}
In further calculations, we'll limit ourselves to the 'quasi-static limit'. This is justified by small size of the contacts $L$ and the sample (no larger than tens of microns) compared to the incident wavelength $\lambda$ (order of 100 $\mu$m). In this approximation, the distribution of current $I(x)$ is parabolic:
\begin{gather}
\label{I-static}
    I(x) \approx i \dfrac{2 \pi E_0 \omega}{\kappa} (x^2-L^2).
\end{gather}
Quite reasonably, the speed of light $c$ does not enter the expressions in the quasi-static limit. In the same limit, it would be more convenient to express the fields in all space from distribution of linear charge density $Q(x)$, not the current. The latter is found with the aid of  continuity equation
\begin{equation}
-i \omega Q(x) + I'(x) = 0.  
\end{equation}
The resulting spatial dependence of charge density is linear,
\begin{equation}
    Q(x) \approx \frac{4 \pi E_0 x}{\kappa}.
\end{equation}
This result is quite intuitive: at large distances, a rod behaves as a small polarizable dipole. At small distances, the field should be singular as one approaches the rod's edge. This field in all space is most conveniently found from the Poisson's equation 
\begin{gather}
\label{phi}
\Delta \varphi = 4 \pi Q(x) \delta(y) \delta(z),
\end{gather}
where we have neglected the finite radius of the wire $a$. Taking the Fourier transform of (\ref{phi}) in the $yz$ plane, we are led to the expression for the components of diffracted electric field in the $z=0$ plane:
\begin{gather}
    \label{E}
  \mathbf{E}_d(\mathbf{q},\omega) = {\bf E}_0 + \mathbf{F}_{\mathbf{q},\omega} E_0,\\
 \mathbf{F}_{\mathbf{q},\omega}  =  \dfrac{16 \pi^2 \mathbf{q}}{ \kappa q} \dfrac{\sin{q_x L} - q_x L \cos{q_x L}}{q_x^2} 
\end{gather}

It is possible to take into account finite width $2 W$ of the contact. We consider a plane contact of nonzero width as a set of parallel perfectly conducting threads, which field was found above.  This modifies the expression for total diffracted field by adding an oscillatory-decaying envelope
\begin{gather}
    \label{W}
    \mathbf{F}_{\mathbf{q},\omega} \rightarrow  \mathbf{F}_{\mathbf{q},\omega} \dfrac{\sin{q_y W}}{q_y W}
\end{gather}

\begin{figure}[h!]
\begin{minipage}[h]{1\linewidth}
\center{\includegraphics[width=0.4\linewidth]{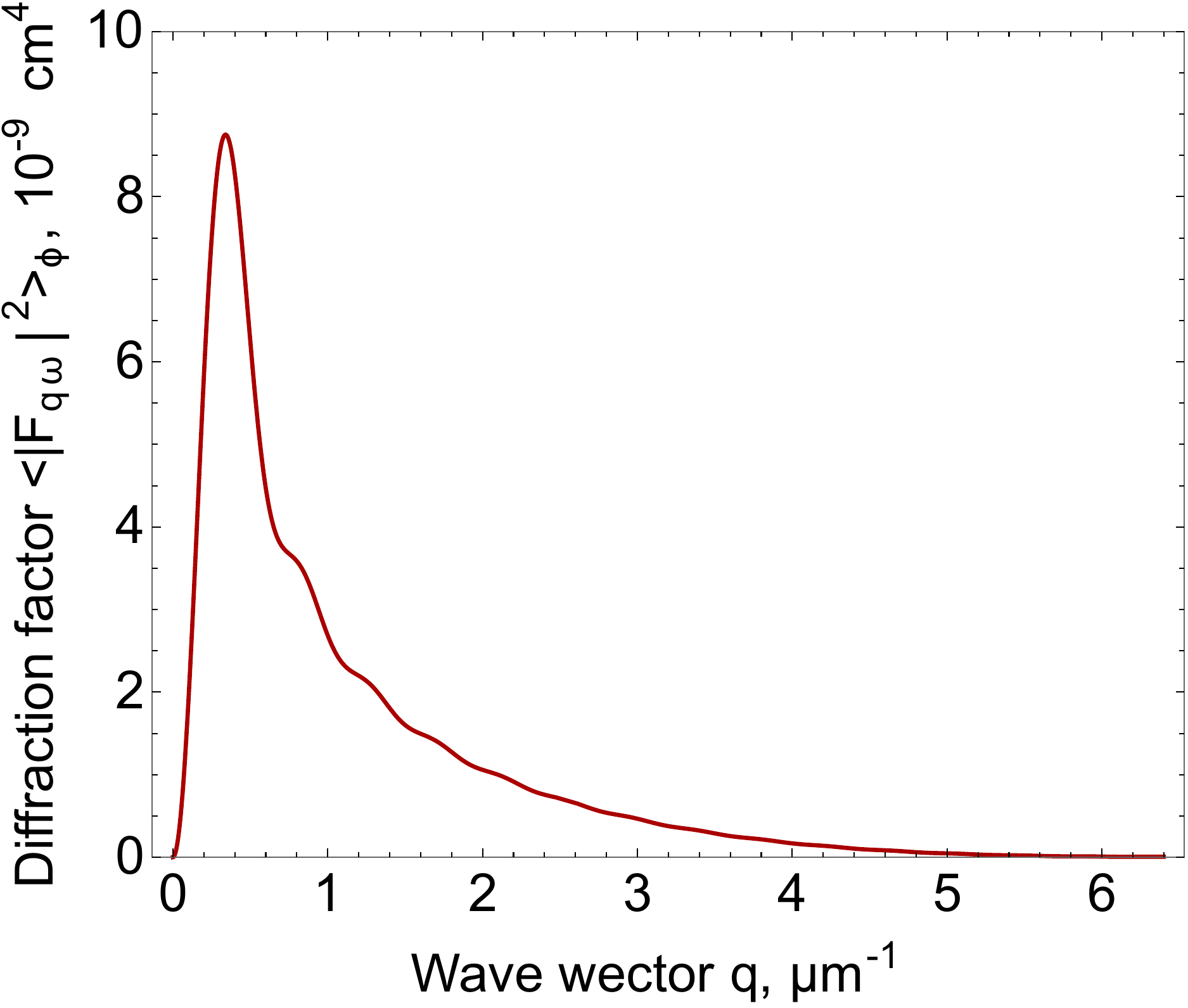} }
\caption{ Angle-averaged diffraction factor $\left< |\mathbf{F}_{\mathbf{q,\omega}}|^2 \right>_{\varphi}$ with contact length 2L = 15 $\mu$m, contact width 2W = 1 $\mu$m and local capacitance $\kappa = 11$ }
\label{Diffraction_amplitude}
\end{minipage}
\end{figure}

The resulting diffraction amplitude $\mathbf{F}_{\mathbf{q},\omega}$ is highly anisotropic. As the conductivity of the 2DES is isotropic, it is convenient to average $\mathbf{F}^2_{\mathbf{q},\omega}$ over the angle $\phi$ in the $xy$-plane. The averaged expression would depend only on modulus of $q$ but not its direction. An example of the angle-averaged diffraction amplitude is shown in Fig.~\ref{Diffraction_amplitude}. It displays a power-law growth at small $q$ corresponding to the power-law decay of the dipole field at large distances. The decay of the diffraction amplitude at large $q$ (small distances) appears mainly due to finite width of the contacts.

\subsubsection{Screening of the diffracted field by electrons in graphene}

The diffracted field ${\bf E}_d$ induced upon diffraction at the leads is further screened (or anti-screened) by the 2DES electrons. This total (self-consistent) field governs the absorption of radiation in 2DES. Strictly speaking, this self-consistent field (not $E_0$) should also act as a source for induced currents in the contacts, eq.~(\ref{eq}). However, if the surface conductivity of 2DES is well below the surface conductivity of the metal lead, $\sigma_{2DES} \ll \sigma_{lead} $, the self-consistency loop can be truncated. We can approximate the net field as $\mathbf{E}(\mathbf{q}, \omega) = \mathbf{E}_d(\mathbf{q}, \omega) / \varepsilon(\mathbf{q},\omega, \omega_c)$, where $\varepsilon(\mathbf{q},\omega, \omega_c)$ is dielectric function of 2DES itself.

A consistent way of obtaining $\varepsilon(\mathbf{q},\omega, \omega_c)$ is to consider the scattering of incident evancescent wave ${\bf E}_d \propto e^{i q_x x + i q_y y} e^{-\sqrt{q^2 - k^2}z}$ by 2DES and underlying substrates, e.g. with the transfer-matrix technique. In the quasi-static limit, a simpler approach is possible. We replace the electric field $\mathbf{E}_d|_{z=0}$ by an equivalent distribution of surface charges $\rho(\mathbf{q})$ in the 2DES plane. Their density is found from the Gauss law:
\begin{equation}
    \mathbf{E}_d|_{z=0} = - 2 \pi \rho(\mathbf{q}) \dfrac{\mathbf{q}}{q}.
\end{equation}
We solve for the electric field created by charges $\rho(\mathbf{q})$ in the presence of 2DES and substrates, and present the result in the form: 
\begin{equation}
\mathbf{E}|_{z=0} =  - \dfrac{2 \pi \rho(\mathbf{q})}{\varepsilon(\mathbf{q},\omega, \omega_c)} \dfrac{\mathbf{q}}{q}.
\end{equation}

In our calculations, we adopt the model of the substrate shown in Fig.~(\ref{geom}). hBN and SiO$_2$ are considered as a joint layer of thickness $d$ with average dielectric constant $\varepsilon \approx 4$. The doped silicon (acting as back gate) is extended to infinity. It has finite bulk conductivity which frequency dependence is assumed in the Drude form:
\begin{equation}
    \sigma_{Si} = \dfrac{\sigma^{dc}_{Si}}{1 - i \omega \tau_{Si}}.
\end{equation}
The equivalent frequency-dependent conductivity of the back gate is given by
\begin{equation}
\varepsilon_{Si}(\omega) = \varepsilon_{Si}(\infty) + \dfrac{4 \pi \sigma_{Si}}{i \omega} 
\end{equation}
where $\varepsilon_{Si}(\infty) \approx 12$ is the high-frequency dielectric constant of silicon.

With these assumptions, is possible to find the effective dielectric constant $\varepsilon$ of the compound structure represented by the graphene layer and the substrate. The net dielectric constant can be factorized into the contributions emerging due to 2D graphene and the substrate, 
\begin{equation}
\varepsilon = \varepsilon_{2D} \varepsilon_{sub}, 
\end{equation}
with individual contribution given by
\begin{gather}
    \label{eps} 
    \varepsilon_{2D}({\bf q}, \omega, \omega_c) = 1 -  \dfrac{2 \pi q \sigma_{xx}(\omega, \omega_c, q)}{i \omega \varepsilon_{sub}({\bf q}, \omega)}, \qquad \varepsilon_{sub}({\bf q}, \omega) = \dfrac{\varepsilon + 1}{2} \dfrac{1 + \dfrac{\varepsilon_{Si} - \varepsilon}{\varepsilon_{Si} + \varepsilon} \dfrac{\varepsilon - 1}{\varepsilon + 1}e^{-2q d} }{1- \dfrac{\varepsilon_{Si} - \varepsilon}{  \varepsilon_{Si} + \varepsilon} e^{-2q d} }
\end{gather}

\begin{figure}[ht!]
\begin{minipage}[h]{1\linewidth}
\center{\includegraphics[width=0.7\linewidth]{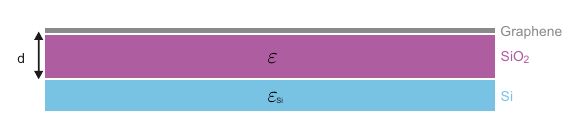} }
\caption{ Structure geometry used in the model. $d = 330$ nm is the distance to the silicon gate. }
\label{geom}
\end{minipage}
\end{figure}

It is possible to find simple limiting expressions for $\varepsilon_{2D}$ and $\varepsilon_{sub}$ in the case of highly conductive silicon gate:

\begin{gather}
    \label{eps} 
    \varepsilon_{2D}({\bf q}, \omega, \omega_c) = 1 -  \dfrac{2 \pi q \sigma_{xx}(\omega, \omega_c, q)}{i \omega \varepsilon_{sub}({\bf q}, \omega)}, \qquad \varepsilon_{sub}({\bf q}, \omega) = \dfrac{\varepsilon + 1}{2} \dfrac{1 +  \dfrac{\varepsilon - 1}{\varepsilon + 1} e^{-2q d} }{1-  e^{-2q d} }
\end{gather}

The simplest expression is obtained in the model of a non-gated structure, when graphene lies on a dielectric layer extending to infinity:

\begin{gather}
    \label{eps} 
    \varepsilon_{2D}({\bf q}, \omega, \omega_c) = 1 -  \dfrac{2 \pi q \sigma_{xx}(\omega, \omega_c, q)}{i \omega \varepsilon_{sub}({\bf q}, \omega)}, \qquad \varepsilon_{sub}({\bf q}, \omega) = \dfrac{\varepsilon + 1}{2}
\end{gather}

\subsubsection{Polarization of diffracted field}

\begin{figure}[h!]
\center{\includegraphics[width=0.5\linewidth]{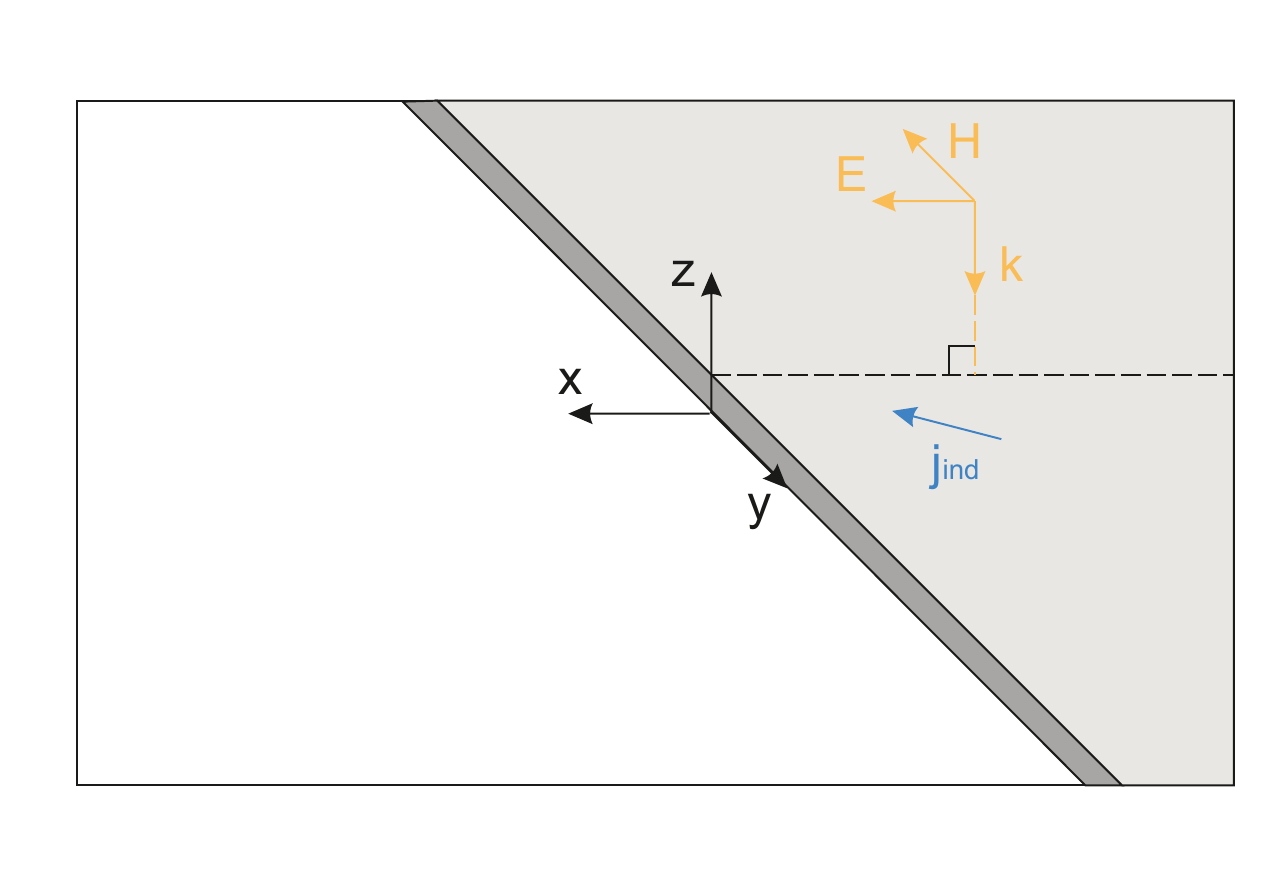} }
\caption{Schematic of electromagnetic wave incident on a half-plane contact}
\label{Half-plane}
\end{figure}

We proceed to demonstrate that the polarization state of the near field is largely governed by the contacts to the 2DES, but not by the polarization state of the incident field. For model contact treated as a thin wire, the $y$-component of the field can not polarize the rod at all, and all near fields are induced by the $x$-component of the incident electric field. One may note that the thin wire model of the contact is not very realistic. We proceed to show that the desired behaviour of polarization persists for another limiting model of a metal contact, a perfectly conducting half-plane (Fig.~\ref{Half-plane}). The boundary of half-plane is directed along the $y$-axis.

The problem of half-plane diffraction was originally discussed by Sommerfeld~\cite{sommerfeld1895}. The distributions of induced sheet current densities ${\bf j}(x)$ can be found by requiring that induced electric field ${\bf E}_d$ exactly compensates the incident field. The induced field themselves can be found from the fundamental solutions of the wave equation (at $z=0$):

\begin{gather}
E_{d x} =-\frac{\pi}{k c}\left(k^{2}+\partial_{x}^{2}\right) \int d x^{\prime} j_{\omega x}(x') H_{0}(k|\Delta x|), \\
E_{d y} =-\frac{\pi}{c} k \int d x^{\prime} j_{\omega y}(x') H_{0}(k|\Delta x|),
\label{Ed}
\end{gather}
where $H_0(x)$ is Hankel function being the fundamental solution of the wave equation, $k = \omega / c$ is the wave vector.

Equations (\ref{Ed}) can be solved by Wiener-Hopf technique~\cite{nobleWiener}. Proceeding in this way, we obtain expressions for the Fourier components of the net field
\begin{equation}
\label{WH-solution}
E_{d x}(q)=\frac{E_{0x}}{i q}\left(\sqrt{1+\frac{q}{k}}-1\right) \qquad
E_{d y}(q)=-\frac{E_{0y}}{i q}\left(1-\frac{1}{\sqrt{1+\frac{q}{k}}}\right).
\end{equation}

Already the Fourier-space solution of the diffraction problem allows us to predict different character of electric fields in the immediate vicinity of the contact. The Fourier harmonics of $x$-component decay slowly at $q \rightarrow \infty$, $E_{dx}(q) \propto q^{-1/2}$. It implies that real-space field is singular at $x\rightarrow -0$, $E_{dx} (x) \propto |x|^{-1/2}$. The situation is different for $y$-component of the field. Its Fourier harmonics decay faster $E_{dy}(q) \propto q^{-1}$. The corresponding real-space field remains finite at $x\rightarrow -0$.




\begin{figure}[h!]
\center{\includegraphics[width=0.5\linewidth]{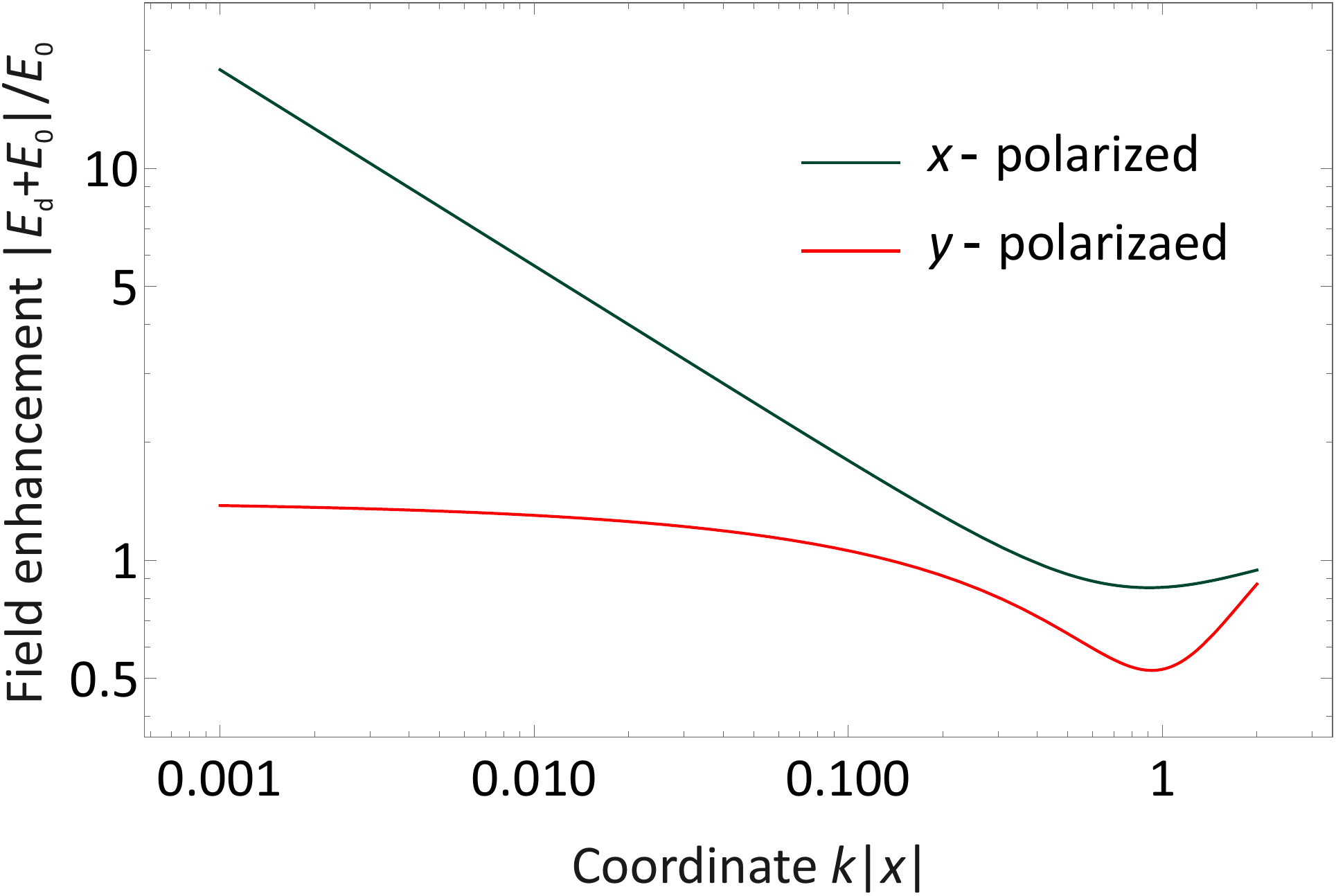} }
\caption{$x$ and $y$ components of the total field generated upon diffraction at the conducting half-plane vs distance from the edge $k|x|$}
\label{Real-space-diffraction}
\end{figure}

The above discussion is fully confirmed by computing the real-space diffracted fields via inverse Fourier transforms of Eq.~(\ref{WH-solution}). The result is shown in Fig.~\ref{Real-space-diffraction}. At distances $k x\sim 10^{-3}$ nm, the $x$-component of the near-field is roughly 10 times stronger, compared to the $y$-component. This finally justifies our assumption that the polarization of the near field is approximately linear, with the dominant component of ${\bf E}$ directed normally to the contacts.

\clearpage
\newpage

\subsection{\textbf{
Nonlocal conductivity of graphene in magnetic field}}
\label{Magnetoabsorption_microscopics}
\subsubsection{Nonlocal conductivity for particle-conserving collisions}

The nonlocal conductivity of 2DES is the key element for evaluation of both absorption and analysis of collective modes in the magnetized 2d electron gas. Following \cite{Chaplik_1985_CR_overtones}, we start with the classical kinetic equation for distribution function of 2d electrons $f$ in constant perpendicular magnetic field ${\bf B}$:
\begin{gather}
	 \dfrac{\partial f}{\partial t} + \mathbf{v} \dfrac{\partial f}{\partial \mathbf{r}} -e \left( \mathbf{E} + \dfrac{1}{c} \mathbf{v} \times \mathbf{B} \right) \dfrac{\partial f}{\partial \mathbf{p}} = - {\rm St}\{f\}.
\end{gather}
Above, ${\bf v} = v_0 {\bf p}/p$ is the velocity of electron with momentum ${\bf p}$, $v_0 \approx 10^6$ m/s is the Fermi velocity in graphene, ${\bf E} \propto e^{i {\bf q r} - i\omega t}$ is the net (self-consistent) field acting on the 2DES, and ${\rm St}\{f\}$ is the collision integral. We proceed to linearize the kinetic equation, $f = f_0 + \delta f e^{-i \omega t + i \mathbf{q}\mathbf{r}}$, where $f_0$ is the equilibrium Fermi function, and $\delta f$ is the small correction proportional to the strength of electric field. The linearized equation reads as:
\begin{equation}
\label{KE}
    - i (\omega - {\bf q v}) \delta f - e  \mathbf{E} \dfrac{\partial f_0}{\partial \mathbf{p}} - \dfrac{1}{c} \mathbf{v} \times \mathbf{B} \dfrac{\partial \delta f}{\partial \mathbf{p}} = - {\rm St}\{\delta f\}.
\end{equation}

The collision integral is adopted in the relaxation-time approximation with a 'restoring term' that acts to conserve the particle number~\cite{Mermin-Lindhard}:
\begin{equation}
    {\rm St}\{\delta f\} = \frac{1}{\tau_{\rm p}} \left(\delta f + \delta \mu \frac{\partial f_0}{\partial \varepsilon}\right).
\end{equation}
Above, $\tau_{\rm p} = \nu^{-1}$ is the transport relaxation time, and $\nu^{-1}$ is the electron collision frequency. The magnitude of 'corrected' Fermi energy $\delta \mu$ is chosen by requiring particle conservation upon collisions:
\begin{equation}
\label{particle-conservation}
    \sum_{\bf p}{{\rm St}\{\delta f\}} = 0.
\end{equation}

Passing to the polar coordinates, we present the kinetic equation (\ref{KE}) in the form: 
	\begin{gather}
	\label{KE_phi}
	 -i(\alpha + i T) \delta f + i \beta \cos{\varphi} \delta f  + \dfrac{\partial \delta f}{\partial \varphi} =  \dfrac{\partial f_0}{\partial p} \left(  \dfrac{e E}{\omega_c}  \cos{\varphi} - \dfrac{\nu \delta \mu}{\omega_c v_0} \right)
	\end{gather}
where $\alpha = \omega / \omega_c$, $\beta = q v_0 / \omega_c$, $T = \nu / \omega_c$, and $w_c$ is the cyclotron frequency. Then, we seek for the function $\delta f$ in the form $\delta f = g e^{-i \beta \sin{\varphi}}$, and expand $g$ as a series of angular harmonics$\varphi$: 
\begin{equation}
 g = \sum g_s e^{i s \varphi}.   
\end{equation}
The solution for $g_s$ reads as:
\begin{equation}
\label{gs}
	 g_s = i \dfrac{\partial f_0}{\partial p} \dfrac{ \frac{e E}{\omega_c}  \frac{s}{\beta}  - \frac{\nu \delta \mu}{\omega_c v_0} }{\alpha + i T - s} J_s(\beta),    
\end{equation}
where $J_s(\beta) = \dfrac{1}{2 \pi} \int_0^{2 \pi} e^{i(\beta \sin{\varphi} - s \varphi)} d \varphi$ is the $s$-th order Bessel function. Apparently, the $s$-th harmonic of the distribution function $g_s$ is excited via $s$-th order cyclotron resonance ($s=1$ corresponds to the main CR). 

The closure of the solution (\ref{gs}) is achieved by fixing $\delta \mu$ from particle number conservation (\ref{particle-conservation}). This results in
\begin{gather}
\label{deltaN}
\delta \mu =  \frac{i v_0 e E}{\omega^2} \dfrac{Y_{00}^{(1)}}{1-\frac{i \nu}{\omega} Y_{00}^{(0)}},
\end{gather}
where we have introduced the dimensionless factors
\begin{equation}
Y_{ij}^{(k)} = \sum_{s=-\infty}^{\infty} \left( \dfrac{s}{\beta}\right)^k \dfrac{ J^{(i)}_s[\beta]J^{(j)}_s[\beta]}{1 - s\omega_c / \omega + i \nu / \omega}
\end{equation}
and $J^{(i)}_s[\beta]$ is the $i$-th order derivative of the Bessel function. 

Equations (\ref{gs}) and (\ref{deltaN}) are sufficient to find the current $j_i = -e \sum_{\bf{p}}v_i \delta f$ and the conductivity. In a practically important limit $\varepsilon_F \gg k T$, we have obtained:
	\begin{gather}
	\label{Sigma}
		 \sigma_{xx}(\omega, q) = 2 \sigma_D \left[ Y_{00}^{(2)} + \dfrac{i \nu}{\omega} \dfrac{(Y_{00}^{(1)})^2}{1-\frac{i \nu}{\omega} Y_{00}^{(0)}}  \right]
		 \qquad
     \sigma_{xy}(\omega, q) = 2 i \sigma_D \left[ Y_{10}^{(1)} + \dfrac{i \nu}{\omega} \dfrac{Y_{00}^{(1)}Y_{10}^{(0)}}{1-\frac{i \nu}{\omega} Y_{00}^{(0)}}  \right],
	\end{gather}
where $\sigma_D = i n e^2/ m_c \omega$ is the collisionless Drude conductivity, $n$ is the  carrier density, and $m_c = p_F / v_0$ is the cyclotron mass.

\subsubsection{Electron-electron collisions and the hydrodynamic limit}
It is instructive to track the evolution of multiple cyclotron resonances with increasing the strength of electron-electron (e-e) collisions $\nu_{ee}$. Contrary to electron-impurity and electron-phonon collisions, e-e collisions conserve the net momentum of colliding particles. To account for this fact, we adopt the e-e collision integral in generalized relaxation-time approximation. The collision integral pushes the distribution function $\delta f$ toward local equilibrium~\cite{Svintsov_PRB_Crossover}
\begin{equation}
\label{EE-collision-integral}
     {\rm St}_{ee}\{\delta f\} = \frac{1}{\tau_{ee}} \left[ \delta f- \frac{\partial f_0}{\partial \varepsilon}\left(\delta \mu + \mathbf{p} \delta \mathbf{u}\right) \right], 
\end{equation}
where $\delta \mu$ and $\delta \mathbf{u}$ are found from the conservation of particle number and momentum: 
\begin{gather}
    \label{ConsLaw}
    \sum_{\mathbf{p}} \left( \delta f - \delta f_{hd} \right) = 0  \qquad
    \sum_{\mathbf{p}} \mathbf{p} \left( \delta f - \delta f_{hd} \right) = 0 
\end{gather}


The kinetic equation with e-e collision integral (\ref{EE-collision-integral}) and conservation laws (\ref{ConsLaw}) are sufficient to describe the behaviour of electron liquid at large wave vectors $q\sim \omega/v_0$ and across the whole hydrodynamic-to-ballistic crossover. We leave the analysis of emerging generalized hydrodynamic equations for further publication. The only fact that we note here is that the poles in conductivity at $\omega = n \omega_c$ are washed out. A more detailed analysis of e-e collision effects on magnetodispersion will be presented in the following section.

Several attempts to treat high-frequency nonlocal conductivity in the magnetic field in the presence of e-e scattering were taken in Refs.~\cite{Polini_Hall_viscosity} and \cite{Alekseev_Magnetic_Viscous_Resonance}. However, none of the obtained results can be valid across the whole hydrodynamic-to-ballistic crossover and at arbitrarily large wave vectors. Ref.~\cite{Alekseev_Magnetic_Viscous_Resonance} attempts to write down the hydrodynamic-type equation valid both at low ($\omega \tau_{ee} \ll 1$) and high ($\omega \tau_{ee} \gg 1$) frequencies. These equations contain spatial derivatives no higher than to the second order, so they naturally do not reproduce the peculiar behaviour of magnetoplasmons at large $q$, including the anticrossing of magnetoplasmons and cyclotron resonances. Ref.~\cite{Polini_Hall_viscosity} expanded the distribution function $\delta f$ into angular harmonics and truncated the expansion at the second order. Naturally, such truncation is not sufficient to describe higher-order cyclotron resonances.

\newpage
\subsection{\textbf{
Theoretical analysis of Bernstein modes and near-field absorption}}
\label{Bernstein_modes}
    
\subsubsection{Bernstein modes}

Collective modes supported by magnetized graphene can be analyzed known the expressions for the dielectric function $\varepsilon_{2D}$ [Eq.~\ref{eps}] and nonlocal conductivity $\sigma({\bf q},\omega)$ [Eq.~\ref{Sigma}]. Complex zeros of $\varepsilon_{2D}$ corresponds to Bernstein modes, which imaginary part governs the wave damping. Here, we refrain from analyzing the complex-valued spectrum of plasmons. Instead, we turn our attention to the so-called 'loss function', or 'plasmon spectral function', ${\rm Im} \varepsilon^{-1}_{2D} ({\bf q}, \omega, \omega_c)$ which is more informative in scattering problems. Both ${\bf q}$ and $\omega$ in this approach are real-valued, while the loss function is complex.

Above, we have shown that ${\rm Im} \varepsilon^{-1}_{2D} ({\bf q}, \omega, \omega_c)$ governs the absorption of ${\bf q}$-th spatial Fourier harmonic of the field. Once real ${\bf q}$ and real $\omega$ satisfy ${\rm Re}\varepsilon_{2D} = 0$, and provided that ${\rm Im}\varepsilon_{2D} \ll 1$, the loss function is resonantly enhanced. In such a situation, the {\it real} frequency of electromagnetic radiation $\omega$ appears close to the complex frequency of (magneto)plasmon mode. Weakly damped plasmon modes are thus seen as peaks in the loss function, which height is inversely proportional to dissipative part of the dielectric function ${\rm Im}\varepsilon_{2D} $.

The dispersion of Bernstein modes is visualized in the color maps of the loss function calculated with Eqs.~(\ref{eps}) and (\ref{Sigma}), Fig.~\ref{Fig_disp}. The Bernstein mode, like an ordinary magnetoplasmon, starts from the cyclotron frequency $\omega_c$. When approaching the doubled cyclotron frequency, it demonstrates a plateau with downward bending. The density of states at the plateau is singular while the group velocity is zero. The frequency of plateau is shifted from the cyclotron overtone by the value of mini-gap~\cite{Volkov2014}:
\begin{equation}
\Delta \sim 10 (a_{\rm B}/R_{\rm c})^2 \omega_{\rm c},  
\end{equation}
where $a_{\rm B}$ is the effective Bohr radius and $R_{\rm c}$ is the cyclotron radius. The wave vector $q^*$ corresponding to the BM plateau can be approximately estimated as crossing point of unperturbed plasmon dispersion $\omega^{(0)}_{\rm pl}(q)$ (evaluated at $B=0$) with $\omega = 2\omega_c$.

The second branch of the Bernstein mode starts from $\omega = 2 \omega_c$ at $q=0$, and forms an anticrossing with the triple cyclotron frequency. Similar anticrossings occur at each multiple of the cyclotron frequency  (Figure~\ref{Fig_disp}, left). Each subsequent mini-gap is smaller than the previous one, and already at $n \sim 4$ the gap becomes comparable with inverse relaxation time. Thus, higher-order resonances are less and less visible in the magnetoabsorption.

Since the frequency of the incident radiation is fixed during measurements, while the magnetic field and the cyclotron frequency are variable, it is more natural to analyze the loss functions in terms of $q$ and $f_c = \omega_c/2\pi$. In these coordinates, the singular point moves to \textit{large} $q$ with increasing \textit{frequency} and to the region of \textit{small} $q$ with increasing \textit{gate voltage}, as shown in (Figure~\ref{Fig_disp}, right).
\begin{figure}[ht!]
\begin{minipage}[h]{1\linewidth}
\center{\includegraphics[width=0.9\linewidth]{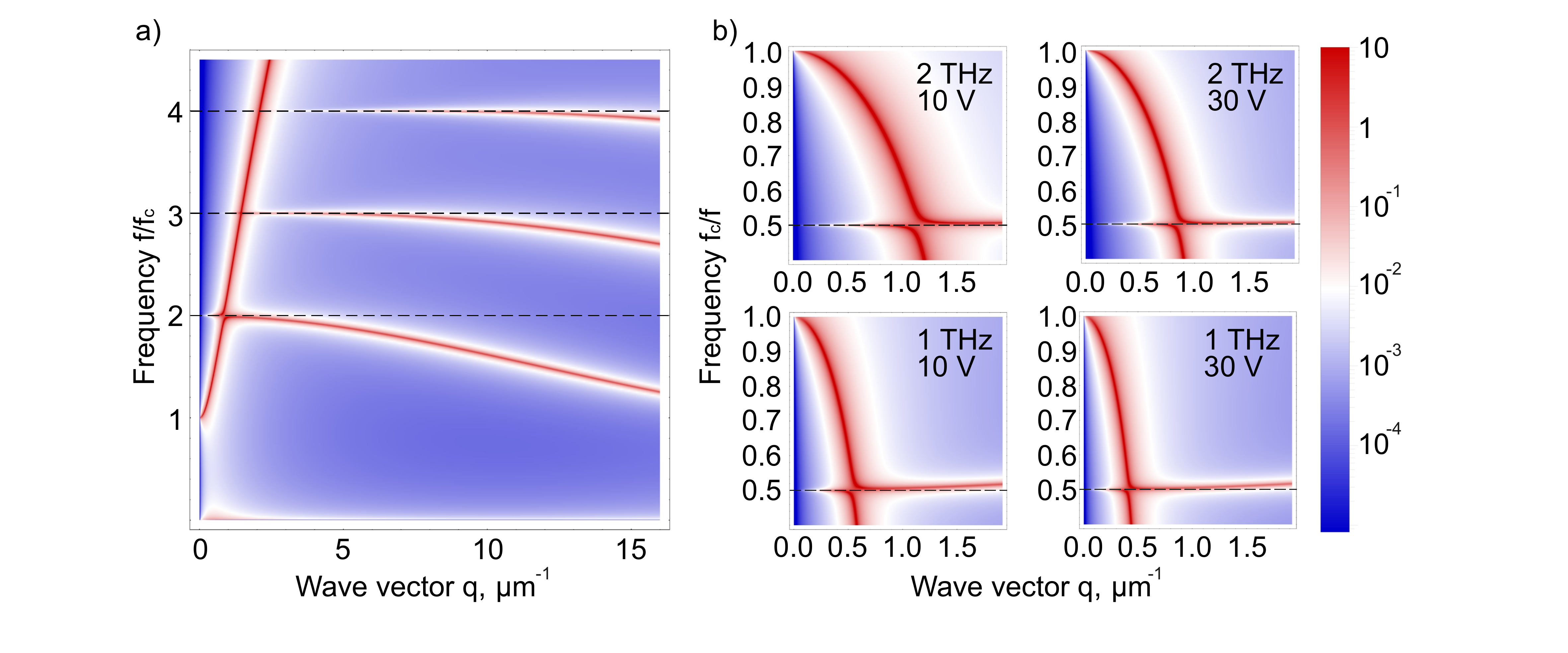} }
\caption{ Magnetoplasmon  dispersion curve visualized through loss function ${\rm Im }\varepsilon^{-1}_{2D}(q,\omega,\omega_c)$ in (a) $(q,\omega)$ coordinates with $f_c = 1$ THz and $V_g = $ 10 V, in (b) $(q,\omega_c)$ coordinates at various gate voltages and frequencies. All curves are obtained at gate distance $d = 330$ nm, $\rho_{Si} = 0$ Ohm$\times$cm, $\varepsilon = 4$. }
\label{Fig_disp}
\end{minipage}
\end{figure}

It is instructive to analyze the smearing of the Bernstein modes for particle- and momentum-conserving e-e collisions. The result of such calculation is shown in Fig.~\ref{BM_and_hydrodynamics}. The anticrossing of magnetoplasmon dispersion with $\omega = 2\omega_c$, while present in the collisionless case, is fully washed out for strong e-e collisions with relaxation time $\tau_{ee} = 100$ fs. No traces of 'transverse magnetosound' mode predicted in Ref.~\cite{Alekseev} are seen in our calculations. We may suggest that the predicted transverse magnetosound mode in highly viscous electron fluid (corresponding to $\omega \tau_{ee} \gg 1$) is equivalent to the second branch of the Bernstein mode starting at $\omega = 2\omega_c$.

\begin{figure}[h!]
\center{\includegraphics[width=1\linewidth]{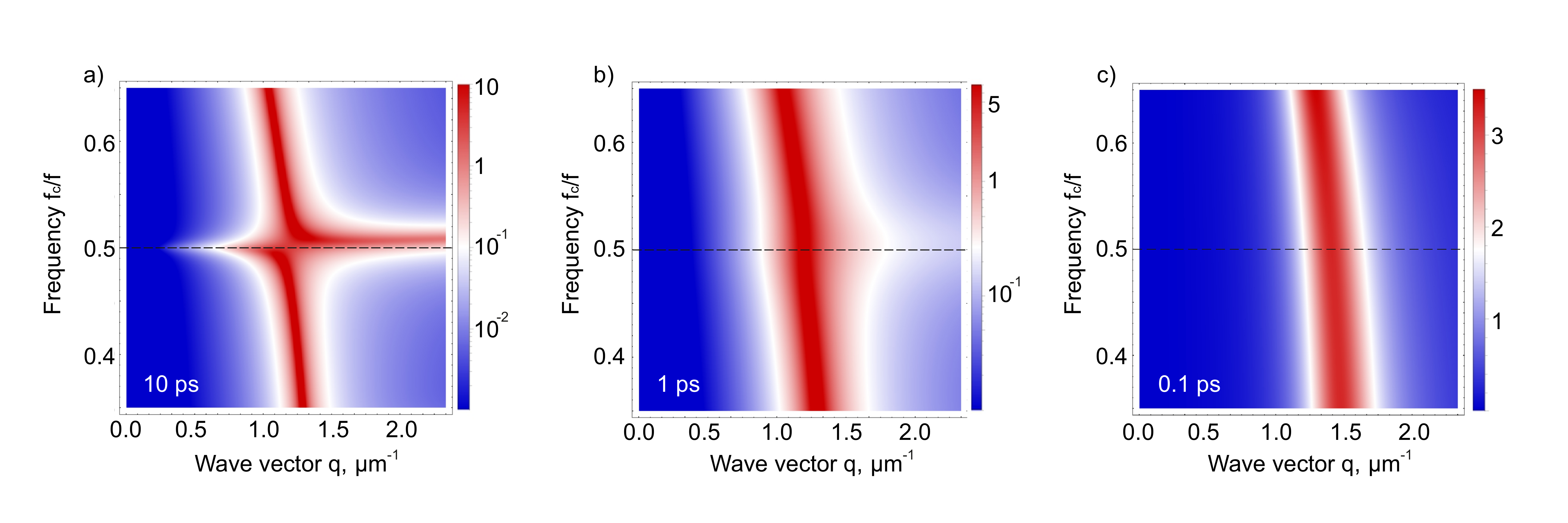} }
\caption{Magnetoplasmon  dispersion curve visualized through loss function ${\rm Im }\varepsilon^{-1}_{2D}(q,\omega,\omega_c)$ in (a) ballistic regime, $\tau_{ee} = 10$ ps, (b) intermediate regime, $\tau_{ee} = 1$ ps, (c) in the hydrodynamic regime $\tau_{ee} = 100$ fs. All curves are obtained at  $V_g = 10V$,  $f = 2$ THz with gate distance $d = 330$ nm, $\rho_{Si} = 0$ Ohm$\times$cm, $\varepsilon$ = 4}
\label{BM_and_hydrodynamics}
\end{figure}
    
\subsubsection{Magnetoabsorption: analytical estimates}
We proceed to single out the contribution of collective magnetoplasmon modes to the absorption. For clarity, we limit ourselves to the ungated case, in which $\varepsilon(\omega, \omega_c, q) = 1 - \frac{2 \pi q}{i \omega} \sigma_{xx}(\omega,\omega_c,q)$. In this case, expression for absorption (\ref{Absorp}) takes on a simple form:
\begin{equation}
P = 2 \int \dfrac{d \mathbf{q}}{(2 \pi)^2} \dfrac{\omega}{2 \pi q} |\mathbf{E}_{\mathbf{q} \omega}|^2 \Im{\left[ \dfrac{1}{\varepsilon(\omega,\omega_c, q)} \right]}.   
\end{equation}

In the vicinity of magnetoplasmon dispersion $q_{mp}(\omega, \omega_c)$, the wave vector dependence of complex dielectric function $\varepsilon = \varepsilon'+ i \varepsilon''$ can be presented as
\begin{equation}
\label{Epsilon_expansion}
    \varepsilon \approx \frac{\partial \varepsilon}{\partial q} [q - q_{\rm mp}(\omega, \omega_c)] + i \varepsilon''.
\end{equation}
Here, $q_{mp}$ is the real wave vector of magnetoplasmon such that $\varepsilon ' (q_{mp},\omega,\omega_c) = 0$. In accordance with expansion (\ref{Epsilon_expansion}), the loss function can be approximated by a delta-function in the integral sense.
\begin{gather}
\label{delta}
    \Im{\left[ \varepsilon^{-1} \right]} = \sum\limits_i{ 
      \frac{\delta (q - q^{(i)}_{mp})}{\left| \partial \varepsilon' / \partial \omega \right| v^{(i)}_{\rm gr}}.
    }
\end{gather}
Above, $v^{(i)}_{\rm gr} = (\partial \omega / \partial q)|_{q = q^{(i)}_{mp}}$ is the magnetoplasmon group velocity. The $i$-summation is performed over all branches of plasmon dispersion at given $\omega$ and $\omega_c$: there exist two branches slightly below each CR overtone, and only a single branch just above $\omega  = n \omega_c$. Representation (\ref{delta}) is justified for low losses, $\varepsilon '' \ll 1$, and, more importantly, not very close to the BM plateau where the group velocity approaches zero. 

Within the approximation of well-defined collective modes, Eq.~(\ref{delta}), the only origin of absorption is the excitation of magnetoplasmons. Their wave vectors $q^{(i)}_{\rm mp}$ are uniquely defined by $\omega$ and $\omega_c$. The delta-function in the expression for absorption is readily integrated, and the net absorbed power is now given by:
\begin{equation}
    P = 2 \frac{\omega}{(2 \pi)^2} 
     \sum\limits_i{
     \frac{\left\langle |\mathbf{E}_{\mathbf{q}\omega}|^2 \right\rangle_{\varphi}}{\left| \partial \varepsilon' / \partial \omega \right| v^{(i)}_{\rm gr}}
     }
\end{equation}
It becomes clear that the absorbed power is inversely proportional to the magnetoplasmon group velocity, which can be interpreted as prolonged interaction of 2d electron system with a slow light. 
    
Naturally, the delta-approximation to the loss function should break down in immediate vicinity of zero group velocity point $(q^*,\omega^*)$. Inclusion of finite losses is mandatory at that point, as well as treatment of plasmon dispersion curvature. We expand the dielectric function as
\begin{equation}
    \varepsilon \approx 
    \left.\frac{\partial \varepsilon'}{\partial \omega}\right|_{q^*\omega^*}
    \left[
    \omega - \omega^* - \frac{1}{2} \left.\frac{\partial^2 \omega_{\rm mp}}{\partial q^2}\right|_{q^*} (q - q^*)^2
    \right] + i \varepsilon''.
\end{equation}
The loss function $\Im{\varepsilon ^{-1}}$ takes the Lorentzian form:
\begin{gather}
    \Im{ \varepsilon^{-1} } \approx \dfrac{\varepsilon''}{\varepsilon''^2 + 
    \left\{ 
    \left.\frac{\partial \varepsilon'}{\partial \omega}\right|_{q^*\omega^*}
    \left[
    \omega - \omega^* - \frac{1}{2} \left.\frac{\partial^2 \omega_{\rm mp}}{\partial q^2}\right|_{q^*} (q - q^*)^2
    \right]
    \right\}^2
    } 
\end{gather}

\begin{figure}[h!]
\begin{minipage}[h]{1\linewidth}
\center{\includegraphics[width=0.5\linewidth]{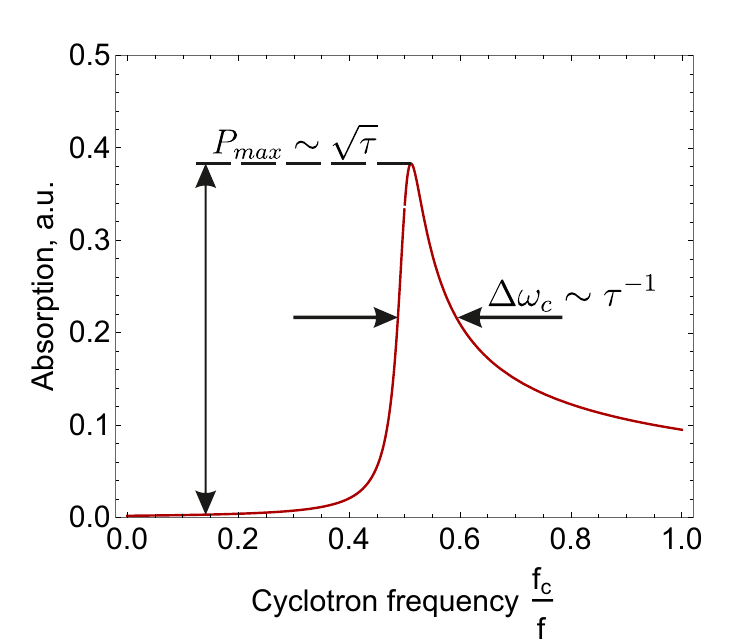} }
\caption{ Plasmon resonance at the doubled cyclotron frequency in the Lorentzian approximation. }
\label{Lorentz}
\end{minipage}
\end{figure}

Assuming the diffraction amplitude to be weakly varying in the vicinity of singular point $q^*$, we obtain an approximate expression for the absorption power:
\begin{gather}
P = \dfrac{\omega}{4 \pi \alpha \sqrt{\gamma}} \Re{\left[  \dfrac{1}{\sqrt{(\omega_c-\omega_0) + i\frac{\delta}{\alpha}}}\right] \left\langle |\mathbf{E}_{\mathbf{q}\omega}|^2_{q=q^*} \right\rangle_{\varphi}}
\end{gather}
where $\delta = \varepsilon''$, $\alpha = \left(\frac{\partial \varepsilon'}{\partial \omega_c} \right)_{q^*\omega^*}$, $\gamma = \left( \frac{\partial^2 \omega_c}{\partial q^2} \right)_{q^*\omega^*}$ are the parameters determined from the dispersion relation magnetoplasmons. Since $\delta$ is proportional to $1/\tau_{\rm p}$, we find that the height of the resonance peak is proportional to the square root of relaxation time, $P_{\max} \propto \sqrt{\tau_{\rm p}}$, and the width is inversely proportional to relaxation time $\Delta \omega_c \sim 1/\tau_{\rm p}$. The resulting shape of absorption curve is shown in Fig.~\ref{Lorentz}.

To find the frequency dependence of the resonance peak, one can roughly estimate the parameters $\alpha \sim V_g \omega^{-3}, \ \delta \sim V_g \tau_{\rm p}^{-1} \omega^{-3}, \ \gamma \sim \omega^{-1} $ in the ungated case. As a result, the scaling of the absorption peak with frequency, gate voltage and relaxation time is given by
\begin{gather}
\label{Pmax}
    P_{\max} \sim  \dfrac{\omega^{9/2}}{V_{\rm g} \tau_{\rm p}^{1/2}} \left\langle |\mathbf{E}_{\mathbf{q}\omega}|^2_{q=q^*} \right\rangle_{\varphi}
\end{gather}
We note that the first factor in Eq.~(\ref{Pmax}) does not fully determine the frequency and gate voltage dependence of absorbed power. Extra dependence can come from diffraction amplitude evaluated at $q=q^*$. Particularly, increase in radiation frequency $\omega$ shifts $q^*$ to larger values, where the diffraction amplitude can be small (see Fig.~\ref{Diffraction_amplitude}). This effect softens the frequency dependence of absorption at the CR overtones. 

\subsubsection{Numerical calculations of magnetoabsorption}

In this section, we proceed to analyze the radiation absorption at CR overtones without restricting ourselves to collective mode approximation. Expressions (\ref{Absorp}), (\ref{Sigma}), (\ref{E}) and (\ref{eps}) form the basis for such analysis. The quantity of interest here is the absorption cross-section $\sigma_{\rm abs}$ which is obtained by dividing the absorbed power (\ref{Absorp}) by the incoming radiation intensity $I = c E_0^2/2\pi$. In all plots, we normalize the absorption cross-section by the sample area $S = W \times L \approx 2.5 \times 10^{-6}$ cm$^2$.

\textit{Gate voltage}. The dependence of absorption cross-section near the cyclotron overtones on gate voltage is shown in Fig.~\ref{Vg}a. All resonances slowly decrease in magnitude with increasing voltage, which is consistent with measurements. One reason for this decrease is increased damping $\varepsilon''$ at larger $V_g$, which is manifested in a pre-factor of Eq.~\ref{Pmax}. Another reason is shift of characteristic wave vector $q^*$ to lower values with increasing $V_g$, which may result in reduced diffraction amplitude $F_{{\bf q},\omega}$.

\begin{figure}[h!]
\center{\includegraphics[width=0.8\linewidth]{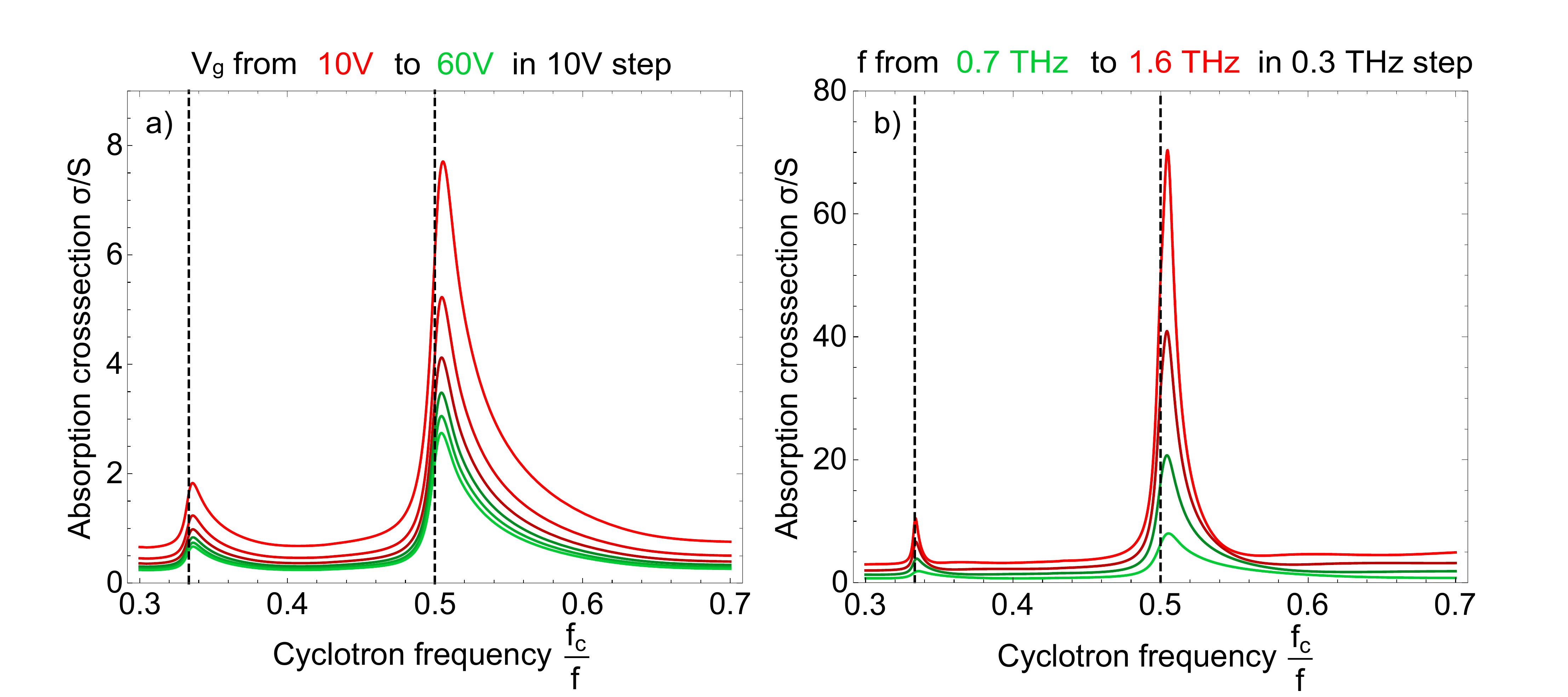} }
\caption{(a) absorption cross section vs cyclotron frequency and gate voltage with 0.69 THz, distance to gate $d =$ 330 nm, relaxation time $\tau_{\rm p} =$ 20 ps, $\rho_{Si} = 0.1$ Ohm*cm and $\tau_{Si} = 0.01$ ps. (b) absorption at the peak resonance point at main CR overtone vs gate voltage and radiation frequency with distance to gate $d =$ 330 nm, relaxation time $\tau_{\rm p} =$ 20 ps, $\rho_{Si}= 0.1$ Ohm*cm and $\tau_{Si} = 0.01$ ps, $\kappa = 11$, $S = 4 \times 10^{-6} $cm$^2$  }
\label{Vg}
\end{figure}

\textit{Radiation frequency}. The maximum absorption cross-section goes up with increasing the radiation frequency, which is illustrated in Fig.~\ref{Vg}b. This trend is quite intuitive: the quality factor $Q = \omega \tau_{\rm p}$ goes up with frequency if the relaxation time is the same. However, this trend is not observed in experiments: the magnitude of absorption at the main overtone is the largest at $f=0.69$ THz, and is roughly the same at $f=1.64$ THz and $f=2.54$ THz. The discrepancy can be attributed to the lack of self-consistency in calculations: screening by graphene electrons may change the magnitude of induced current in the leads, which was not accounted for in our calculations.

\textit{Silicon substrate resistivity}. By varying the dc resistivity of Si substrate $\rho_{\rm Si}$, one can interpolate in theory between the cases of gated graphene ($\rho_{\rm Si} \rightarrow 0 $) and ungated graphene ($\rho_{\rm Si}\rightarrow \infty$). With increasing the conductivity of the silicon, the absorption at magnetoplasmonic modes is enhanced and a characteristic non-resonant absorption background is seen (Fig.~\ref{rho_silicon}). In the case of low conductivity, the background is suppressed and CR overtones acquire high contrast. Enhancement of near-field absorption in a gated structure ($\rho_{\rm Si} \rightarrow 0$) can be attributed to overall softening of plasmon modes and reduction of their group velocity. 

The behavior observed in the measurements is more consistent with the case of high resistivity of Si substrate, $\rho_{Si} > 0.01$ $\Omega\cdot$cm. The nominal dc resistivity of Si substrate is $0.001...0.005$ $\Omega\cdot$cm at room temperature. At the same time, an order of magnitude increase in dc resistivity is possible at experimental temperatures $T \sim 4$ K.

\begin{figure}[h!]
\center{\includegraphics[width=0.45\linewidth]{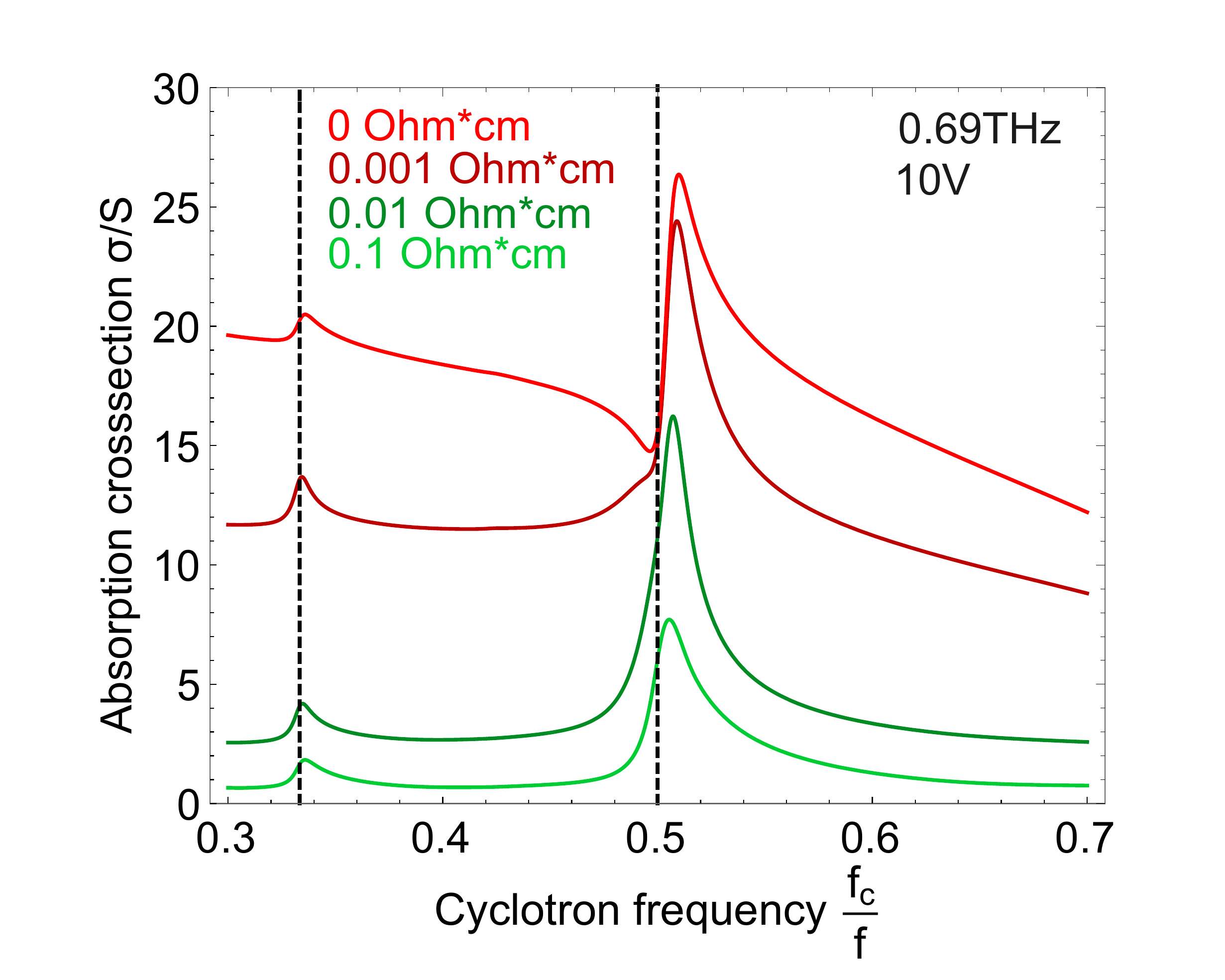} }
\caption{Absorption cross section vs cyclotron frequency and silicon resistivity with 0.69 THz, $V_g = 10$V, distance to gate $d =$ 330 nm, relaxation time $\tau_{\rm p} =$ 20 ps, $\tau_{Si} = 0.01$ ps, $\kappa = 11$, $S = 4 \times 10^{-6} $cm$^2$ }
\label{rho_silicon}
\end{figure}

\textit{Relaxation time}. With an increase in the relaxation time, the quality factor of the resonance naturally increases, as shown in Fig.~\ref{tau}. As the BM resonance is associated with singular plasmonic density of states, its dependence on $\tau_{\rm p}$ is not as sharp as that for ordinary CR (see Eq.~\ref{Pmax}). In particular, increase in relaxation time from 10 ps to 30 ps results in $\sim 50$ \% enhancement of resonance visibility. 

\begin{figure}[h!]
\center{\includegraphics[width=0.45\linewidth]{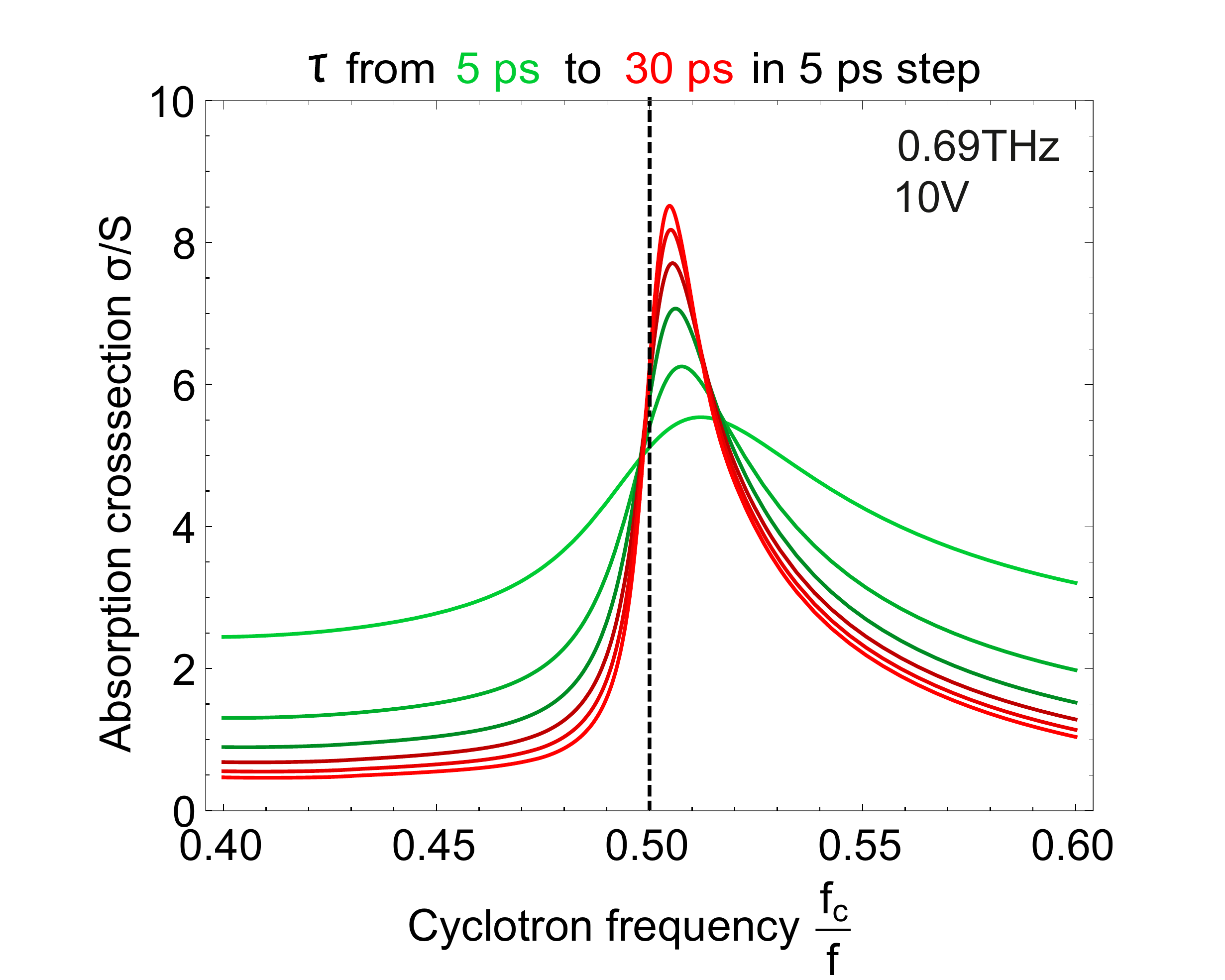} }
\caption{Left: absorption cross section vs cyclotron frequency and relaxation time with $f =$ 0.69 THz, $V_g = 10$V distance to gate $d = 330$ nm and silicon resistivity $\rho_{Si} = $0.1 Ohm*cm, $\tau_{Si} = 0.01$ ps, $\kappa = 11$, $S = 4 \times 10^{-6} $cm$^2$. Right: absorption at the peak resonance point of main CR overtone vs relaxation time and gate voltage with 0.69 THz, distance to gate $d = 330$ nm and silicon resistivity $\rho_{Si} = $0.1 Ohm*cm, $\tau_{Si} = 0.01$ ps, $\kappa = 11$, $S = 4 \times 10^{-6} $cm$^2$.}
\label{tau}
\end{figure}



\clearpage
\newpage

\subsection{\textbf{
Radiative decay of the main cyclotron resonance}}
\label{Radiative_decay}

The main cyclotron resonance is naturally sensitive to the helicity of radiation: absorption of circularly polarized light is possible for only one direction of $B$. To calculate the absorption at the main CR, we supplement the Joule's law (\ref{Absorp}) with the transfer-matrix technique used to relate the fields in 2DES plane $\{E_x,E_y\}$ with the incident field $\{E_{0x},E_{0y}\}$. In a practical case $d \ll \lambda$, this relation simplifies to
\begin{gather}
\label{Polarization_rotation}
     \begin{pmatrix}
                E_x \\ E_y
    \end{pmatrix} = \dfrac{1}{\Delta} \begin{pmatrix}
                \frac{1+n_{Si}}{2} + \frac{2 \pi}{c} \sigma_{xx} & \frac{2 \pi}{c} \sigma_{xy}  \\ - \frac{2 \pi}{c} \sigma_{xy} & \frac{1+n_{Si}}{2} + \frac{2 \pi}{c} \sigma_{xx}
    \end{pmatrix}  \begin{pmatrix}
                E_{0x} \\ E_{0y}
    \end{pmatrix} ,\\
    \Delta = \left(\frac{1+n_{Si}}{2} + \frac{2 \pi}{c} \sigma_{xx} \right)^2 + \left( \frac{2 \pi}{c} \sigma_{xy} \right)^2
\end{gather}

We further introduce the dimensionless absorption coefficient $\alpha$ being the ratio of absorbed power density and the incoming light intensity. Known the fields in 2DES plane $\{E_x,E_y\}$, the absorption coefficient is evaluated as:
\begin{gather}
    \alpha = \dfrac{4 \pi}{c} \left( \sigma'_{xx} \dfrac{|E_x|^2+|E_y|^2}{|E_{0x}|^2 + |E_{0y}|^2} + i \sigma''_{xy} \dfrac{E_x E^*_y - E^*_x E_y}{|E_{0x}|^2 + |E_{0y}|^2} \right)
\end{gather}
Evaluation of absorption based on Eqs.~(\ref{Polarization_rotation}) and (\ref{Abs_main_CR})  show that the main CR is indeed helicity-sensitive (Fig.~\ref{MainCR}, left). With increasing gate voltage, the magnitude of resonance quickly goes down. The effect is nothing but radiative decay of CR. In other terms, it can be attributed to the screening of incident field by graphene itself. In a practically interesting case $\omega_c \tau_{\rm p} \gg 1$ the absorption coefficient in the vicinity of the resonant frequency can be presented as

\begin{gather}
\label{Abs_main_CR}
     \alpha = \dfrac{2}{1+n'_{Si}} \dfrac{\tau_p \tau'_r}{(\tau_p + \tau'_r)^2} \dfrac{1}{1 + (\omega - \dfrac{1}{\tau''_r} - \omega_c)^2 \tau_{\Sigma}^2} \left(1 +  \dfrac{2 \Im{\left[E_{0x} E^*_{0y}\right]}}{|E_{0x}|^2 + |E_{0y}|^2}  \right), \ \  \ \ \ \ 
\end{gather}
where $\tau_{\Sigma}^{-1} = \tau_p^{-1} + \tau_r^{-1}$ is the effective decay rate, $\tau^{-1}_r = \dfrac{2}{1+n_{Si}} \dfrac{2 \pi \sigma_0}{c \tau_p} $ is the complex radiative decay rate (the imaginary part corresponds to the shift of the resonant frequency), $\sigma_0 = n e^2 \tau_p/m_c$ is the 2D static conductivity. It is apparent from Eq.~\ref{Abs_main_CR} that the main CR is degrading with increasing carrier density due to enhanced $\sigma_0$ and $\tau_r$.

\begin{figure}[h!]
\center{\includegraphics[width=0.8\linewidth]{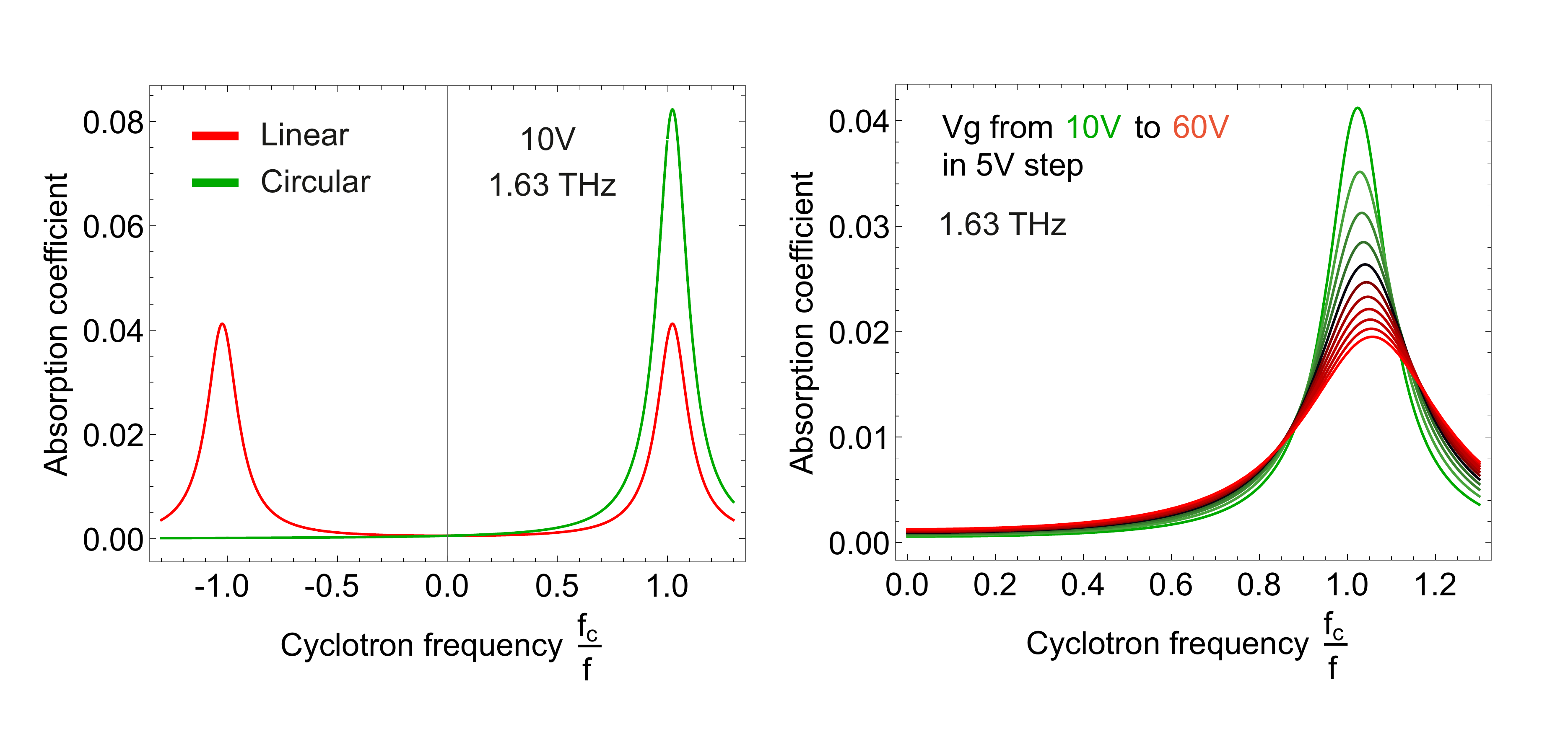} }
\caption{Left: Polarisation dependence of main cyclotron resonance with $V_g = 10$ V. Right: Voltage dependence of main cyclotron resonance. Silicon resistivity $\rho_{Si} = 0.1$ Ohm*cm, silicon relaxation time $\tau_{Si} = 0.01$ ps, 2DES relaxation time $\tau_{\rm p} = 10$ ps.  }
\label{MainCR}
\end{figure}

\clearpage
\newpage

\subsection{\textbf{
Embedded versus remote contacts }}

The key role in the excitation of Bernstein modes, similar to other collective modes in two dimensions, should be played by the inhomogeneity of electric fields due to the presence of sharp contacts. To verify the idea, we have fabricated an additional device of comparable quality but contacted by metal leads outside the main channel.  We found strong photoresponse in the vicinity of the CR harmonics only for invasive case, while for distant leads the photoresponse was dominated by THz-induced magnetooscillations~\cite{Monch2020}.

\begin{figure*}[ht!]
	\centering\includegraphics[width=0.8\linewidth]{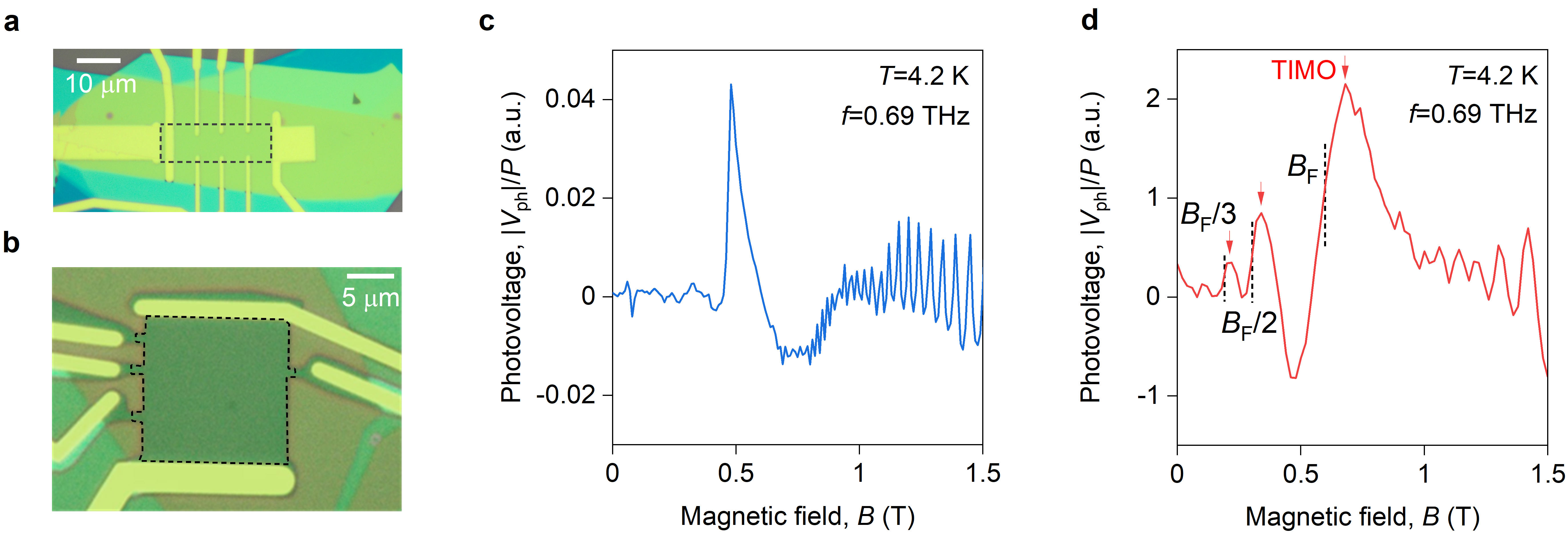}
	\caption{\textbf{CR overtones and TIMO.} \textbf{a-b,} Optical photograph of the device with embedded (a) and remote contacts (b). \textbf{c,} Photovoltage $V_\mathrm{ph}$ as a function of magnetic field $B$ measured in the device from (a) in response to incident  $f=0.69~$THz radiation. \textbf{d,} $V_\mathrm{ph}$ versus $B$ for the same $f$ but measured in the device with remote gold contacts.   }
	\label{FigS7}
\end{figure*}

\clearpage
\newpage





\end{widetext}

\end{document}